\UseRawInputEncoding
\documentclass[reprint,amsmath,amssymb,aps,prb]{revtex4-2}
\pdfoutput=1

\usepackage[colorlinks,citecolor=blue,linkcolor=blue]{hyperref}
\usepackage{graphicx}
\usepackage{epstopdf}
\usepackage{amsmath,bm}
\usepackage{threeparttable}
\usepackage{multirow}
\usepackage{booktabs}
\usepackage{color}
\usepackage{gensymb}
\usepackage{longtable}
 \usepackage{multirow}
\usepackage{xr-hyper}
\usepackage{chngcntr}
\usepackage{appendix}
\usepackage{amsmath}
\usepackage[strings]{underscore}
\usepackage{CJKutf8}

\begin{document}
\begin{CJK}{UTF8}{gbsn}

\title{Thermal and mechanical properties and the structural phase transition under pressure in $A$In$_{2}$As$_{2}$ ($A$=Ca, Sr, Ba)}%
\author{Wen-Ti Guo$^{1,2}$}
\author{Zhigao Huang$^{1,2}$}%
\author{Jian-Min Zhang$^{1,2}$}%
\email[Corresponding author]{jmzhang@fjnu.edu.cn}
\affiliation{1 Fujian Provincial Key Laboratory of Quantum Manipulation and New Energy Materials, College of Physics and Energy, Fujian Normal University, Fuzhou 350117, China
}
\affiliation{2 Fujian Provincial Collaborative Innovation Center for Advanced High-Field Superconducting Materials and Engineering, Fuzhou, 350117, China
}

\begin{abstract}
Experimental results that BaIn$_{2}$As$_{2}$ and Ca(Sr)In$_{2}$As$_{2}$, which are the same class of alkali metal compounds, belong to different structural phases have puzzled the current materials physics community. Here, we investigate the pressure-induced structural phase transition of $A$In$_{2}$As$_{2}$ and its accompanying improvement in mechanical and thermal properties. Firstly, the structural stability of the materials and their structural phase transitions under pressure are characterised by enthalpy and double-checking by phonon dispersion spectrum. We also confirm the structural phase transitions of the hexagonal and monoclinic phases from a group-theoretic point of view, associating their symmetry operations using transformation matrices. In terms of mechanical properties, we propose an effective scheme for pressure modulation of the anisotropy of $A$In$_{2}$As$_{2}$ materials and to induce the transformation of $A$In$_{2}$As$_{2}$ from isotropic to anisotropic (hexagonal) and from brittle to ductile (hexagonal and monoclinic). Meanwhile, we find the negative Poisson's ratio phenomenon under compression and tension, which is favorable for a wide range of applications of this series of materials in aerospace, medicine, sensors, etc. In terms of thermal properties, applying pressure will enhance the structural phase transition temperature of $A$In$_{2}$As$_{2}$ materials to near room temperature. We further give direct evidence of phonon softening based on group velocity calculations and reveal that phonon softening prevents the heat capacity from reaching the Dulong-Petit limit. Our study provides a theoretical basis for selecting stable structural phases and pioneering thermodynamic property studies of the thermoelectric topological candidate material $A$In$_{2}$As$_{2}$.
\end{abstract}

\maketitle

\section{Introduction}
EuX$_{2}$As$_{2}$ ($X$ = Cd, In, Sn), as a series of topologically magnetic material, has been of great interest to topological thermoelectric community because of the intrinsically novel properties.  The layered structure of the Zintl-Klemm phase, EuSn$_{2}$As$_{2}$, can be easily peeled off \cite{c4ref12}. Both theoretical and experimental-based studies have demonstrated that it is an intrinsically magnetic topological insulator \cite{c5ref13}. Other related studies applied high pressure modulation of EuSn$_{2}$As$_{2}$ material to achieve a continuous transition from \textit{R}$\bar{3}$\textit{m} phase to \textit{C}2/\textit{m} phase \cite{c6ref14}, and to the high-pressure rhombohedral phase \cite{PhysRevB.104.L220101}. In addition, EuCd$_{2}$As$_{2}$ is considered to be a Dirac semimetal \cite{b1ref1,b2ref2,b5ref3}. It will undergo a topological phase transition by reforming the magnetic moment direction under the action of pressure  \cite{b1ref1} or an electric field \cite{b4ref4}. Similarly, as an intrinsic magnetic topological insulator, EuIn$_{2}$As$_{2}$ has higher-order topological insulator and axion insulator features \cite{a3ref6,a4ref7,a7ref8,a8ref9,a9ref10}. It also exists magnetic configuration-dependent topological phase transition\cite{a8ref9,a9ref10, a11ref11}.

Alkaline earth ($A$) metal substituted Eu positions will achieve rich non-magnetic topological states, which are reflected in both Sr(Ba)Cd$_{2}$As$_{2}$\cite{b8ref15,c1ref16,SHAH2020121589} and SrSn$_{2}$As$_{2}$\cite{c2ref17,c3ref18}. Likewise, our previous paper reported that AIn2As2 (A =
Ca, Sr, Ba) can achieve both metal-insulator phase transitions and topological quantum phase transitions under the action of pressure\cite{D2CP01764D}. Meanwhile, CaIn$_{2}$As$_{2}$ and SrIn$_{2}$As$_{2}$ have been reported to have a \textit{P}6$_{3}$/\textit{mmc} phase, while BaIn$_{2}$As$_{2}$ possesses a \textit{P}2/\textit{m} phase\cite{newref19}. Why do compounds of Ba-, also an alkaline earth metal, behave in a different phase to compounds of Ca- and Sr-? This is a key question that needs to be urgently explored. Perhaps there are structural phase transitions between them? If there exists structural phase transition, what is the pattern? Are there other intrinsic physical properties that might accompany them? These are the crucial questions that have plagued the experimental field and the reasons that have stimulated research into them in the field of theoretical computing.

As a key means of experimentally regulating the physical properties of materials, pressure is also a research method of particular interest for theoretical calculations. Pressure often induces interesting and important potential properties in materials. For example, a topological phase transition will be achieved by applying hydrostatic pressure in MnBi$_{2}$Te$_{4}$\cite{Cui66401,Pressure-induced} and Cd$_{3}$As$_{2}$\cite{PhysRevB.102.195110,PhysRevB.104.155140}. The pressure will also obtain a metal-insulator phase transition accompanied by a change in the band gap\cite{ruan2019band-structure,Chen2019,ActaPhysicaSinica.70.047101}. In general, the application of pressure inevitably results in structural phase changes. Hydrostatic pressure modulation of the MnBi$_{4}$Te$_{7}$ appears as a structural phase transition\cite{Cui66401}. Tensile and compressive strains lead to multiple phase transitions in photovoltaic films Cs$M$I$_{3}$ ($M$ = Pb, Sn)\cite{C9RA10791F}. Related studies have reported that narrow bandgap Sr$X$$_{2}$As$_{2}$ ($X$=Cd, Sn) materials are easily modulated by external fields\cite{c1ref16,c2ref17}. Pressure is an effective means of studying the properties and relationships between the different structural phases of a material.

Here, we apply pressure to $A$In$_{2}$As$_{2}$ with three different structural phases (\textit{P}6$_{3}$/\textit{mmc}, \textit{R}$\bar{3}$\textit{m}, and \textit{P}2/\textit{m}) to achieve materials rich in structural phase transition. CaIn$_{2}$As$_{2}$ and SrIn$_{2}$As$_{2}$ exhibit a \textit{P}6$_{3}$/\textit{mmc} or \textit{R}$\bar{3}$\textit{m} phase at low pressure, which transforms into a \textit{P}2/\textit{m} phase with increasing pressure. However, BaIn$_{2}$As$_{2}$ tends to form a \textit{P}2/\textit{m} phase. The phases of the three materials are in agreement with the reported experimental results\cite{newref19}. The stability of these phases is next determined by phonon spectral calculations. And we further analyzed the change of the phonon irreducible representation of the pressure-induced structural phase transition. Furthermore, we have investigated the effect of hydrostatic pressure on the mechanical and thermodynamic properties of the material. Our paper explains the physical properties of the structural phase differences between BaIn$_{2}$As$_{2}$ and CaIn$_{2}$As$_{2}$ (SrIn$_{2}$As$_{2}$) and assesses their structural stability and thermal and mechanical characteristics.
\begin{figure}[!htbp]
	\begin{centering}
		\includegraphics[width=0.4\textwidth]{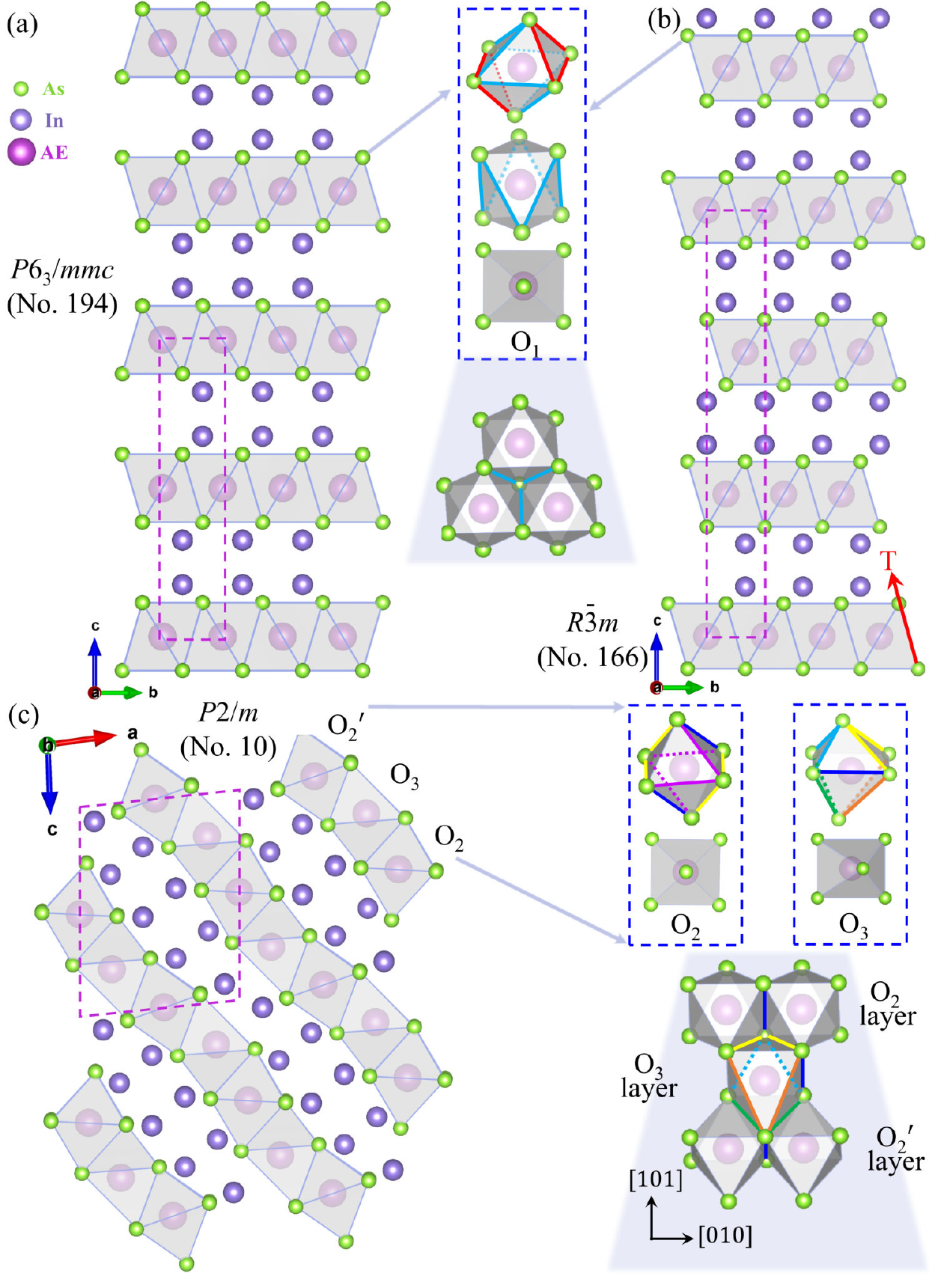}
		\par\end{centering}
	\caption{$A$In$_{2}$As$_{2}$ crystal structure with (a) \textit{P}6$_{3}$/\textit{mmc}, (b) \textit{R}$\bar{3}$\textit{m}, and (c) \textit{P}2/\textit{m} space groups. A schematic diagram of the $A$-As octahedra forming the corresponding structures is also included.}
	\label{fig1}
\end{figure}

A brief synopsis of the subsequent content of this paper is given here. Section \ref{sub2} presents the crystal structures and detailed calculations of DFT.  Section \ref{sec3} focuses on the results of the study and discussion. Sub-sub-section \ref{sub3-1-1} investigates the energy and lattice structure characterization at different pressures and proposes structural phase transitions. Sub-sub-section \ref{sub3-1-2} investigates the structural phase transitions in pressure-modulated systems employing phonon spectroscopy. Sub-sub-section \ref{sub3-1-3} explains the physical nature of structural phase transitions utilizing symmetry shifts in point groups. Sub-section \ref{sub3-2} focuses on the mechanical and thermal properties of materials under pressure modulation. Sub-sub-section \ref{sub3-2-1} analyses the crystalline anisotropy of the material. Sub-sub-section \ref{sub3-2-2} characterizes the material's thermal properties and discusses the realization of negative Poisson's ratio (NPR) performance modulation in compression and tension. Sub-sub-section \ref{sub3-2-3} characterizes the material's thermal properties and reports the pressure-boosted $A$In$_{2}$As$_{2}$ structural phase transition temperature. Sub-sub-section \ref{sub3-2-4} reveals the phenomenon of zero group velocity (ZGV) induced by softening of phonon modes under pressure and the enhancement of thermal conductivity. Section \ref{sec4} provides a summary of this study. Appendix \ref{secA} presents the details of the remaining auxiliary calculation methods. Appendix \ref{subA-1} gives information on the calculation of phonon dispersion spectra and thermodynamic parameters. Appendix \ref{subA-2} presents the relevant parameters for the characterization of mechanical properties, including equations for the calculation of elastic modulus, mechanical stability criterion, crystal anisotropy, calculation of chemical bonding information, and hardness analysis. In the Appendix \ref{secB}, we present and discuss other complementary results, such as the evolution of lattice parameters, symmetry transformation of point group, elastic modulus analysis, chemical bonding, toughness and brittleness, and hardness prediction. Supplemental Material (SM) \cite{SI} gives additional figures related to mechanical and thermal properties.

\section{\label{sub2}Calculation Methods and Crystal Structures}
First-principles calculations based on density functional theory (DFT) are performed in Vienna \textit{ab initio} simulation package (VASP)\cite{PhysRevB.48.13115,PhysRevB.54.11169} based on projected augmented wave (PAW)\cite{PhysRevB.50.17953} and Perdew-Burke-Ernzerhof (PBE) type generalized gradient approximation (GGA)\cite{PhysRevLett.77.3865} exchange-correlation function. The valance wave functions are expanded on plane-wave basis with a 400 eV energy cutoff. In addition, the \textit{s} semi-core orbital of the $A$ atoms are considered as a valence electron. Spin-orbit coupling (SOC) was considered in all our calculations. For ion relaxation, the absolute magnitude of the force on each atom is reduced to less than 0.02 eV/\AA.
For $A$In$_{2}$As$_{2}$ with three kinds of space groups [\textit{P}6$_{3}$/\textit{mmc} (No. 194), \textit{R}$\bar{3}$\textit{m} (No. 166) and \textit{P}2/\textit{m} (No. 10)], the $\Gamma$-centered Monkhorst-Pack \textit{k}-point mesh is considered as 11$\times$11$\times$3, 21$\times$21$\times$3, and 5$\times$13$\times$4, respectively.

According to report, the structures of CaIn$_{2}$As$_{2}$ and SrIn$_{2}$As$_{2}$ are crystallized as EuIn$_{2}$P$_{2}$ type with \textit{P}6$_{3}$/\textit{mmc} phase, whereas BaIn$_{2}$As$_{2}$ is crystallized in the monoclinic EuGa$_{2}$P$_{2}$ structure type with \textit{P}2/\textit{m} phase\cite{newref19}. We further found that the hexagonal structure with \textit{R}$\bar{3}$\textit{m} phase may also exist in these materials. To understand this structural phase difference, pressure was used to systematically study $A$In$_{2}$As$_{2}$ ($A$ = Ca, Sr, Ba) for three space groups.

Three structures are obtained by arranging octahedral structural units and In atoms in different ways. For \textit{P}6$_{3}$/\textit{mmc} phase, $A$ atom occupies the $2a$ position while In and As occupy the $4f$ position. The adjacent octahedral lattices of the \textit{P}6$_{3}$/\textit{mmc} phase, which was labeled as $O_{1}$, form a mirror-symmetric alternating stack between them in the $z$-direction, and two In atomic layers are inserted in between, also mirror-symmetric about the $z$-direction. The right part of Fig. \ref{fig1}(a) gives a schematic diagram of the $O_{1}$ octahedral lattice, with the isosceles triangular planes marked in cyan color and the equilateral triangular planes depicted in red color. The Ba-As octahedral of both \textit{P}6$_{3}$/\textit{mmc} and \textit{R}$\bar{3}$\textit{m} phases are connected by the edges in cyan color in the schematic diagram. For \textit{R}$\bar{3}$\textit{m} phase, $A$ atom occupies the $3a$ position while In and As occupy the $6c$ position. The In atomic layers of the \textit{R}$\bar{3}$\textit{m} phase are arranged similarly to \textit{P}6$_{3}$/\textit{mmc} phase, while the octahedral structural units are arranged along a translational stacking in the $T$ direction as shown by the red arrow in Fig. \ref{fig1}(b).

However, the structure with \textit{P}2/\textit{m} space group with low symmetry is quite different from the previous two. The Ba-As octahedral layer of the \textit{P}2/\textit{m} phase consists of a combination of two types of octahedral lattices, $O_{2}$ and $O_{3}$, as shown in the right part of Fig. \ref{fig1}(c). $O_{2}$ consists of four isosceles triangles (marked in purple and yellow colors) and four irregular triangles. $O_{3}$ consists of four isosceles triangles with different edge lengths, indicated by different colors. Compared to $O_{1}$ and $O_{2}$, the top view of $O_{3}$ has a clear shift of the As and $A$ atoms. As shown in the shaded background part of the octahedral schematic in Fig. \ref{fig1}(c), the [101] orientation, one end of the $O_{3}$ octahedral lattice is connected to the yellow-colored edge labeled in $O_{2}$ through the yellow-colored edge, which is noted as $O_{2}$-$O_{3}$. And the other end of $O_{3}$ is co-edged with the yellow-colored part of $O_{2}$ through the labeled green color edge, which is noted as $O_{2}'$-$O_{3}$. Note in particular that although in the [101] orientation both ends of $O_{3}$ are connected to the yellow-colored part of $O_{2}$ by co-edges, their edge lengths are not equal and depend on the lengths of the yellow-colored and green-colored parts of $O_{3}$, respectively. In the [010] orientation, both $O_{2}$ and $O_{2}'$ layers are spliced via the blue-colored co-edge in the $O_{2}$ octahedral lattice, and the $O_{3}$ layer is also connected by the blue-colored edge of the $O_{2}$ octahedral lattice.

Figs. \ref{fig1}(a), \ref{fig1}(b) and \ref{fig1}(c) show the structures of $A$In$_{2}$As$_{2}$, which belongs to the space group \textit{P}6$_{3}$/\textit{mmc}, \textit{R}$\bar{3}$\textit{m} and \textit{P}2/\textit{m}, respectively. The hexagonal structure of the \textit{P}6$_{3}$/\textit{mmc} and \textit{R}$\bar{3}$\textit{m} phases is composed of alternating [In$_{2}$As$_{2}$]$^{2-}$ layers separated by a slab of $A$$^{2+}$ cations. The structure of \textit{P}2/\textit{m} phase is also layered and it is composed of different types of polyanions [In$_{2}$As$_{2}$]$^{2-}$ units and $A$$^{2+}$ cations. They all exist structural units formed by octahedral with $A$ atoms at the center and As atoms occupying the vertices. The specific structural distinctions are described accordingly in the SM. In short, the valence electron numbers of all three compounds follow the Zintl-Klemm formalism and all elements achieve closed-shell electronic configurations. Lattice parameters reported experimentally are \textit{a} = 4.148 (4.222) {\AA} and \textit{c} = 17.726 (18.110) {\AA} for CaIn$_{2}$As$_{2}$ (SrIn$_{2}$As$_{2}$) with \textit{P}6$_{3}$/\textit{mmc} space group and \textit{a} = 10.275 {\AA}, \textit{b} = 4.301 {\AA}, \textit{c} = 13.332 {\AA}, and \textit{$\beta$} = 95.569 degree for \textit{P}2/\textit{m} space group.

The symmetry generators of \textit{P}6$_{3}$/\textit{mmc} contain identity operation $\mathcal{E}$, inversion symmetry $\mathcal{I}$, twofold screw rotation axis $\mathcal{G}_{2z} = \{C_{2z}|00\frac{1}{2}\}$, threefold rotation axis $C_{3z}$, and the combined rotation axis $C_{2(110)}$. Slightly different with \textit{P}6$_{3}$/\textit{mmc}, \textit{R}$\bar{3}$\textit{m} space group with a hexagonal lattice lacks the $\mathcal{G}_{2z}$ operation but has an additional lattice translation operation $T  = \{x+\frac{2}{3}, y+\frac{1}{3},z+\frac{1}{3}\}$. While the \textit{P}2/\textit{m} has lower symmetry generators that named twofold screw rotation axis $C_{2y}$ (unique axis \textit{b}), identity operation $\mathcal{E}$, and inversion symmetry $\mathcal{I}$.
These basic operations will generate a total of 24, 36 (12$\times$3 sets), and four symmetric operations for the \textit{P}6$_{3}$/\textit{mmc}, \textit{R}$\bar{3}$\textit{m} and \textit{P}2/\textit{m} space groups, respectively.

\section{\label{sec3}Results and Discussion}
\subsection{\label{sub3-1} Structural Stability and Structural Phase Transition}
\subsubsection{\label{sub3-1-1} Dependence of enthalpy on pressure in different structural phases}
Enthalpy is an important state parameter in thermodynamics that characterizes the energy of a material system. It is equal to the sum of the product of internal energy and pressure and volume and can be expressed as,
$H = U + pV$, where $U$ is the internal energy of the system, $p$ is the pressure of the system, and $V$ is the volume.

\begin{figure*}[!htbp]
	\begin{centering}
		\includegraphics[width=1\textwidth]{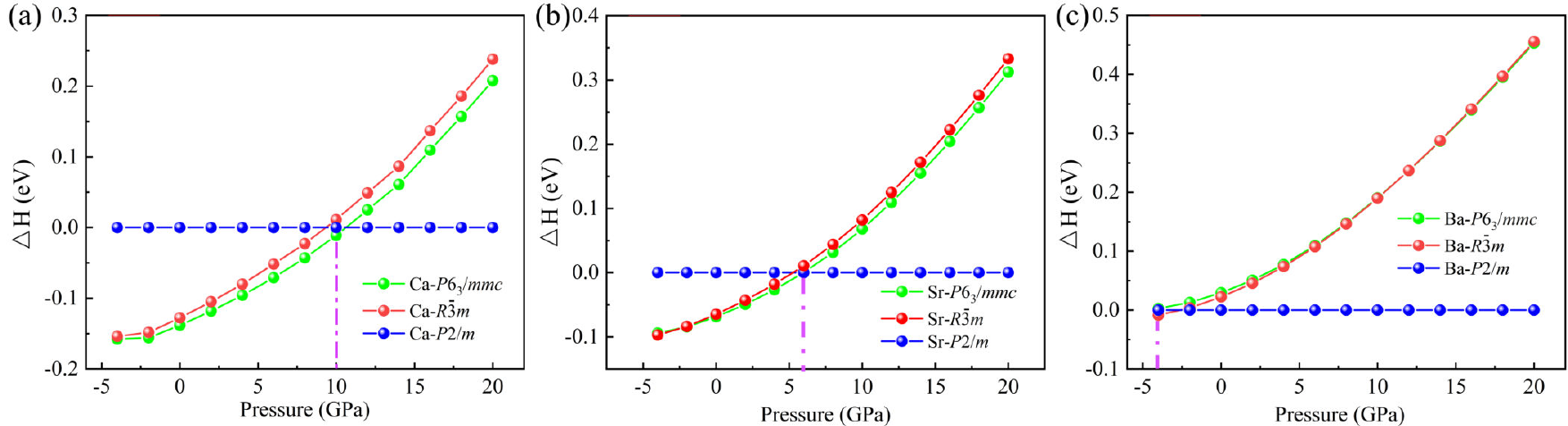}
		\par\end{centering}
	\centering{}\caption{The enthalpy differences of the three structured phases of (a) CaIn$_{2}$As$_{2}$, (b) SrIn$_{2}$As$_{2}$, and (c) BaIn$_{2}$As$_{2}$, with the energy of the \textit{P}2/\textit{m} phase as the reference value.}
	\label{fig2}
\end{figure*}

Thus, we first investigated the enthalpy of different structural phases of $A$In$_{2}$As$_{2}$ under the controling of pressure [see Fig. \ref{fig2}(a)-(c)]. In different $A$In$_{2}$As$_{2}$ systems, the enthalpy difference ($\Delta H$) between the two hexagonal phases (\textit{P}6$_{3}$/\textit{mmc} and \textit{R}$\bar{3}$\textit{m}) under pressure relative to the monoclinic phase (\textit{P}2/\textit{m}) has different trends. Since \textit{P}6$_{3}$/\textit{mmc} and \textit{R}$\bar{3}$\textit{m} have similar crystal structures and symmetry operations, their pressure-dependent enthalpy evolution trends behave approximately the same [see green and red curves in Fig. \ref{fig2}(a)-(c)]. The purple dashed lines in Figs. \ref{fig2}(a)-(c) mark the approximate values of the transition pressure of the structural phase transition, with the left side of the transition point indicating a more likely formation of the hexagonal phase (\textit{P}6$_{3}$/\textit{mmc} or \textit{R}$\bar{3}$\textit{m}), while the right region indicates a more likely formation of the monoclinic phase with \textit{P}2/\textit{m} space group. For CaIn$_{2}$As$_{2}$, SrIn$_{2}$As$_{2}$ and BaIn$_{2}$As$_{2}$ systems, the phase transition points move toward low pressure, respectively, and BaIn$_{2}$As$_{2}$ in particular basically tends to exhibit a \textit{P}2/\textit{m} phase, which is consistent with the experimentally reported results\cite{newref19}. As summarized in Table \ref{tabA1}(d), unlike BaIn$_{2}$As$_{2}$, CaIn$_{2}$As$_{2}$ and SrIn$_{2}$As$_{2}$ tend to form hexagonal structured phases at pressures below 10 GPa and 6 GPa, which explains the experimental conclusion that CaIn$_{2}$As$_{2}$ and SrIn$_{2}$As$_{2}$ have a different space group structure than BaIn$_{2}$As$_{2}$. For BaIn$_{2}$As$_{2}$, a negative-pressure mixed phase (NPMP) with similar energy of the three structural phases will appear at tension stress (negative pressure values), while a \textit{P}2/\textit{m} high-pressure phase (HPP) will formed at compressive stress (positive pressure values).
Thus, we achieved a series of pressure-dependent structural phase transitions for the $A$In$_{2}$As$_{2}$ systems. The change in hardness of the hexagonal and monoclinic phases under pressure is predicted from Table \ref{tabA1}. See Appendix \ref{subB-3-4} for details.

\begin{table}
\centering
\caption{Summary table of $A$In$_{2}$As$_{2}$ different structural phases dependent on
pressure.($+$) indicates a relatively stable structural phase and ($-$) indicates a relatively unstable structural phase.
NPMP and HPP are abbreviations for negative-pressure mixed phase and high-pressure phase, respectively.
}
\begin{tabular}{cccc}
\hline
\hline
Space group                  & CaIn$_{2}$As$_{2}$                            & SrIn$_{2}$As$_{2}$                          & BaIn$_{2}$As$_{2}$                   \\
\hline
\textit{P}6$_{3}$/\textit{mmc} & $\leq $ 10 GPa($+$) & $\textless$ 6 GPa ($+$) & $\textless$ 0 GPa (NPMP)              \\
\textit{R}$\bar{3}$\textit{m}  &      ($-$)      & $\textless$ 6 GPa($+$)  &$\textless$0 GPa (NPMP)              \\
\multirow{2}{*}{\textit{P}2/\textit{m}}  & \multirow{2}{*}{$\textgreater$ 10 GPa($+$)} & \multirow{2}{*}{$\geq $ 6 GPa($+$)}  & $\textless$ 0GPa (NPMP), \\
&　　　&　　　&　　$\geq $ 0 GPa (HPP)　\\
\hline
\hline
\end{tabular}
\label{tab1}
\end{table}

\subsubsection{\label{sub3-1-2}Phonon dispersion spectrum analysis}

\begin{figure*}[!htbp]
	\begin{centering}
		\includegraphics[width=0.8\textwidth]{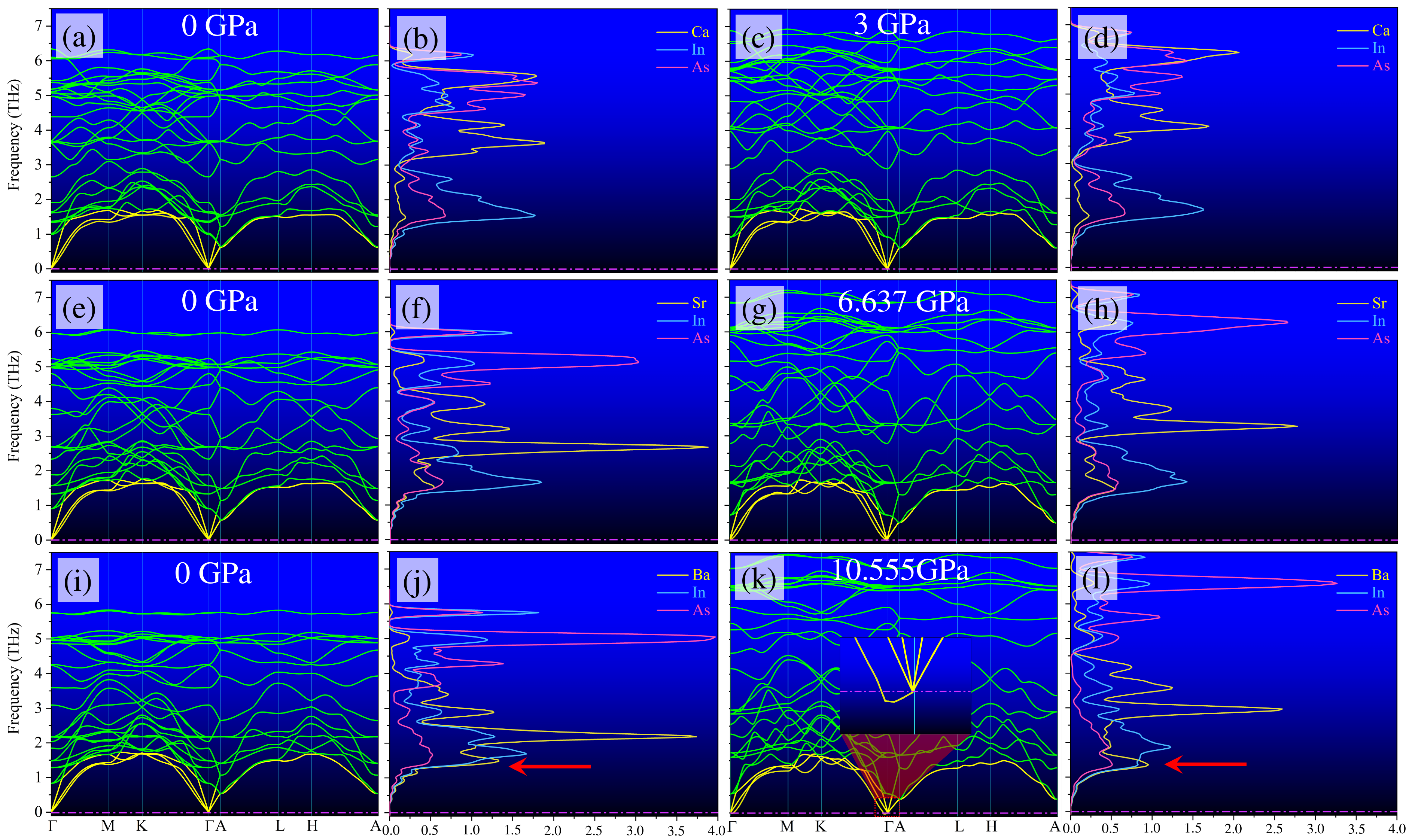}
		\par\end{centering}
	\centering{}\caption{Phonon dispersion and local phonon density of states combinations for (a)-(d) CaIn$_{2}$As$_{2}$, (e)-(h) SrIn$_{2}$As$_{2}$, and (i)-(l) BaIn$_{2}$As$_{2}$ with \textit{P}6$_{3}$/\textit{mmc} space group at atmospheric pressure and zero-bandgap pressure, respectively.}
	\label{fig3}
\end{figure*}

To better illustrate the structural phase transition of $A$In$_{2}$As$_{2}$, we further compare the structural stability of $A$In$_{2}$As$_{2}$ under pressure for different space groups by phonon dispersion spectroscopy calculations.
As shown in Fig. \ref{fig3}, the phonon and projected density of states (PDOS) calculations show that the $A$In$_{2}$As$_{2}$ systems of the \textit{P}6$_{3}$/\textit{mmc} space group are all stable structures at both atmospheric pressure and zero-bandgap pressure. The pressure values at which the zero band gap appears in the induced system have been reported in previous study and are 3 GPa, 6.637 GPa, and 10.555 GPa for CaIn$_{2}$As$_{2}$, SrIn$_{2}$As$_{2}$, and BaIn$_{2}$As$_{2}$, respectively\cite{D2CP01764D}. And we have shown that the system will undergo a non-trivial to trivial topological transitions at these pressure critical values\cite{D2CP01764D}. The lattice waves of the acoustic and optical branches are distinguished in the phonon spectrum by yellow and green curves, respectively. From Figs. \ref{fig3}(a)-\ref{fig3}(l), the acoustic branching lattice waves of the systems with space group \textit{P}6$_{3}$/\textit{mmc} have complete degeneracy in the A-L, L-H, and H-A high-symmetry paths. The BaIn$_{2}$As$_{2}$ of the \textit{R}$\bar{3}$\textit{m} and \textit{P}2/\textit{m} space groups don't have fully phonon dispersion degeneracy in any of the Brillouin zone paths we have considered (see Fig. S11 within the SM\cite{SI}). 

From the PDOS images in Fig. \ref{fig3}, it can be found that the low-frequency parts of CaIn$_{2}$As$_{2}$ and SrIn$_{2}$As$_{2}$ are mainly composed of the phonon dispersion of the In element, while the contribution of the Ba element in the BaIn$_{2}$As$_{2}$ systems are more prominent in the low-frequency phonon dispersion, as the red arrow shown in Figs. \ref{fig3}(j) and \ref{fig3}(l). Two relatively flat high-frequency phonon dispersions consisting of As and In elements exist for the SrIn$_{2}$As$_{2}$ and BaIn$_{2}$As$_{2}$ systems, corresponding to the local peaks in the PDOS diagrams. Similarly, there is a local peak in the 2-3 THz region consisting mainly of $A$ elements. Compared to the atmospheric pressure system, the phonon spectrum of the zero-bandgap pressure system has a broader distribution and spreads to the high-frequency region (see Fig. \ref{fig3}). For BaIn$_{2}$As$_{2}$ structures with different space groups, all have stable phonon characteristics at atmospheric pressure have shown in Figs. \ref{fig3} and S11\cite{SI}. As shown in Fig. S11(b) within the SM\cite{SI}, the acoustic branching lattice wave has a slight imaginary frequency near $\Gamma$, indicating that the structure with \textit{R}$\bar{3}$\textit{m} space group is less stable under the action of 14 GPa than at atmospheric pressure. From Fig. S11(d) within the SM\cite{SI}, the structure with \textit{P}2/\textit{m} space group can still exist stably when 14 GPa is applied. Thus, for the BaIn$_{2}$As$_{2}$ system, as shown in the enlarged plots in Figs. \ref{fig3}(k), and S11(b)\cite{SI}, and S11(d)\cite{SI}, it is shown that the hexagonal phase is not stable at high pressure, while the monoclinic phase with the \textit{P}2/\textit{m} space group is stable. The phonon spectrum calculation verifies our result that BaIn$_{2}$As$_{2}$ has a NPMP and a HPP (\textit{P}2/\textit{m} space group), as considered from the energy comparison.
In a word, our high-pressure calculations realized the structural phase transition of $A$In$_{2}$As$_{2}$ bulk materials. And we further reveal that they are pressure-tunable and can exist stably in a specific pressure range, which is beneficial for the experimentally study.

\subsubsection{\label{sub3-1-3}Symmetry transformation of point group}
\begin{figure*}[!htbp]
	\begin{centering}
		\includegraphics[width=0.8\textwidth]{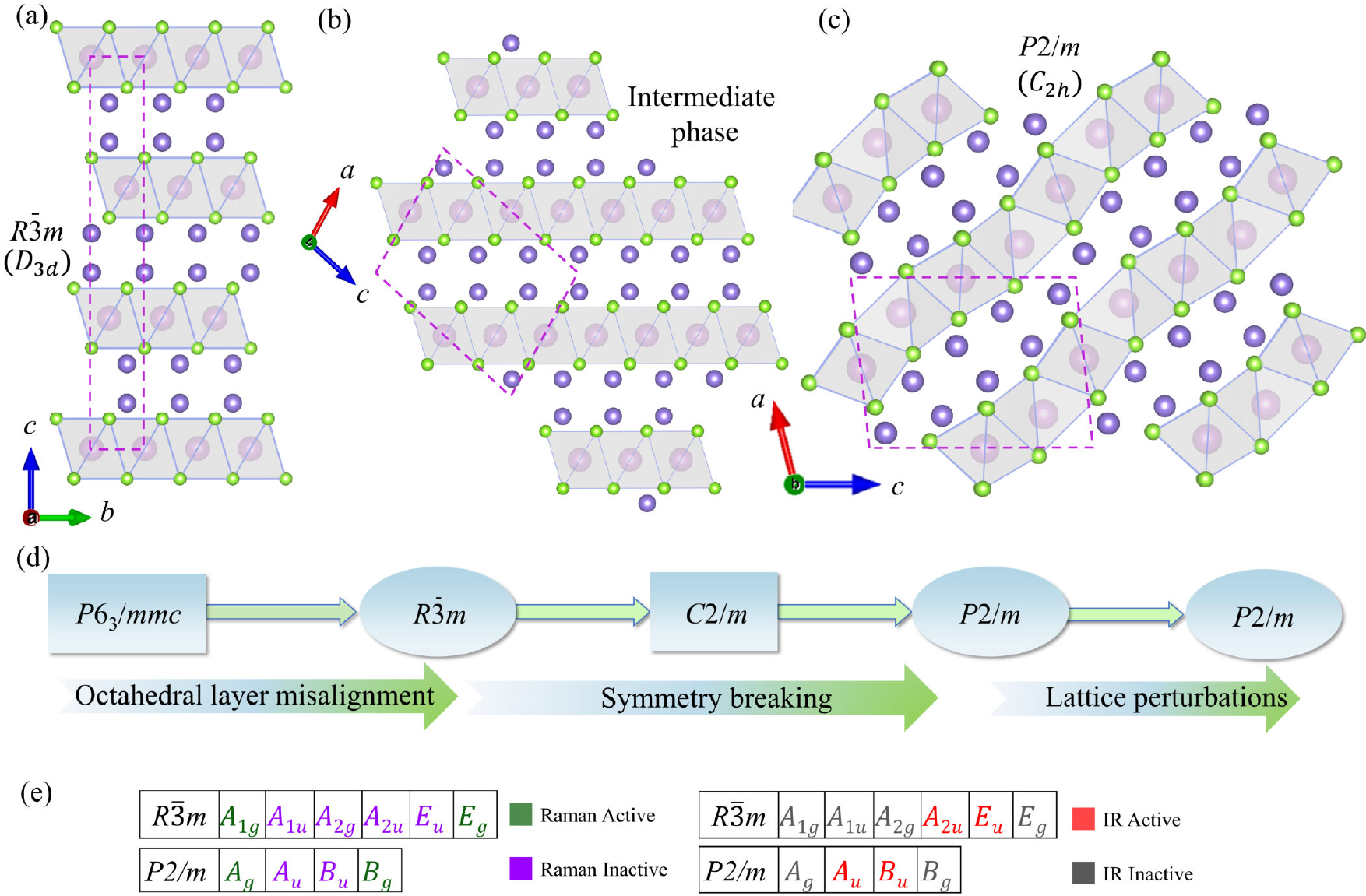}
		\par\end{centering}
	\centering{}\caption{ (a)-(c) Crystal structure of the \textit{R}$\bar{3}$\textit{m} phase, the crystal structure of the intermediate phase \textit{P}${2}$/\textit{m} obtained by a transformation matrix, and our calculated crystal structure of the monoclinic phase (\textit{P}${2}$/\textit{m}). (d) Schematic diagram of the transition from hexagonal to monoclinic phase with the reason for the structural phase transition labeled below. (e) Raman and infrared Raman (IR) activity characterization of \textit{R}$\bar{3}$\textit{m} and \textit{P}${2}$/\textit{m} irreducible representation. }
	\label{fig4}
\end{figure*}

Here, we perform a detailed symmetry theory analysis of the structural phase transition. The structure of the fully relaxed \textit{P}6$_{3}$/\textit{mmc} space group cannot be directly establish a symmetry transition with the \textit{P}2/\textit{m}. But we note that the \textit{P}6$_{3}$/\textit{mmc} structure belongs to the same hexagonal crystal system as \textit{R}$\bar{3}$\textit{m}. Their lattice structures are very similar and only a simple lattice perturbation is required to achieve the structural transformation. Then, the \textit{R}$\bar{3}$\textit{m} phase structure can be transformed into the \textit{P}2/\textit{m} phase through a series of symmetry transformations, as shown in Fig. \ref{fig4}(d).
To understand more deeply the evolutionary mechanism behind the structural phase transition under pressure, we calculate the Raman and Infrared-Raman (IR) activity for these two space groups [see Fig. \ref{fig4}(e) and Table \ref{tabB1}]. The phonon modes at $\Gamma$ point can be decomposed into different irreducible representations, and the correspondence between the irreducible representations of the two phases is shown in Table \ref{tabB2}.

We utilize the overall transformation matrix $T$ [Eq. (\ref{equ18})] to realize the structural phase transition from \textit{R}$\bar{3}$\textit{m} to \textit{P}${2}$/\textit{m} [from Figs. \ref{fig4}(a) to \ref{fig4}(b)], and then the lattice perturbation to obtain Fig. \ref{fig4}(c). $T$ can be obtained by $GS\times EAN \times EEN$, and $LC = GS \times EAN$, where the transformation matrices group-subgroup ,element of the affine normalizers，lattice compatible, and element of the euclidean normalizers are represented by $GS$, $EAN$, $LC$, and $EEN$, respectively. As shown in Table \ref{tabB2}, the irreducible representations of the two point groups at $\Gamma$ have a clear correspondence. It is worth noting that A$_{1u}$ and A$_{2g}$ of \textit{R}$\bar{3}$\textit{m} are both Raman inactive and IR inactive [see Fig. \ref{fig4}(e)]. On the other hand, determining exactly which atoms contribute to these activities will be one of the most important factors influencing the trend of the structural phase transition. As shown in Table \ref{tabB1}, the Raman activity A$_{g}$ (B$_{g}$) of the \textit{P}${2}$/\textit{m} phase is mainly in the $m$ site symmetry group, which can be contributed by In, As, or $A$ atoms at the $2n$ Wyckoff site. And the IR-active A$_{u}$(B$_{u}$) can be contributed by any site of atoms. For the \textit{R}$\bar{3}$\textit{m} phase, the Raman activities A$_{1g}$ and E$_{g}$ are contributed by In, As atoms only, while the IR activities A$_{2u}$ and E$_{u}$ can be contributed by any kind of atoms as well. 

Furthermore, we note the existence of an intermediate phase $C2m$ for this phase change process. A total of six transformation matrix channels with indices [3 2 2] are available for the conversion of the symmetric operation between these two phases, as shown in Eq. (\ref{equ19}). The result obtained by their structure relations of group $G=\textit{R}\bar{3}\textit{m}$ and sub-group $H=\textit{P}{2}/\textit{m}$ belongs to a class with the chain \textit{R}$\bar{3}$\textit{m} $\rightarrow$ \textit{C}${2}$/\textit{m} $\rightarrow$ \textit{P}${2}$/\textit{m} $\rightarrow$ \textit{P}${2}$/\textit{m} and index 12. To change the basis of the group general positions is used the transformation matrices $P=(P,p)$ are shown in Eq. (\ref{equ19}). The linear part $P_{i}$ of the transformation $P=(P,p)$ implies the change of basis vectors, and the column $p$ describes the origin shift $O'= O + p$. And the symmetric operations of group \textit{R}$\bar{3}$\textit{m} (see Table \ref{tabB3}) and subgroup \textit{P}${2}$/\textit{m} (see Table \ref{tabB4}) can be fully correlated by $R_{\textit{P}{2}/\textit{m}} = Q \times R_{\textit{R}\bar{3}\textit{m}} \times P$, where $Q$ is the inverse transformation of $P$. 

According to this relationship, the identity ($\varepsilon$) and inverse ($I$) symmetry operations with low symmetry can naturally be represented by the corresponding ones with high symmetry. However, $C_{2y}$ and $IC_{2y}$ in the \textit{P}${2}$/\textit{m} phase can have different \textit{R}$\bar{3}$\textit{m} transition symmetry operations, which are $C_{2x}$, $IC_{2x}$ ($P_{1}$ and $P_{2}$) or $C_{2y}$, $IC_{2y}$ ($P_{3}$ and $P_{4}$) or $C_{2xy}$, $IC_{2xy}$ ($P_{5}$ and $P_{6}$), respectively. For example, the following Eq. (\ref{equ19}) gives the $C_{2y}$ symmetric operation of the \textit{P}${2}$/\textit{m} phase based on the $P_{1}$ transformation matrix using the $C_{2x}$ symmetric operation of the \textit{R}$\bar{3}$\textit{m} phase. 

In conclusion, we achieved the structural phase transition of the $A$In$_{2}$As$_{2}$ system from the hexagonal phase (\textit{P}6$_{3}$/\textit{mmc} and \textit{R}$\bar{3}$\textit{m}) to the monoclinic phase (\textit{P}${2}$/\textit{m}) from the symmetry operation point of view. As summarized by the schematic diagram of the structural phase transition in Fig. \ref{fig4}(d), the \textit{P}6$_{3}$/\textit{mmc} phase can be transformed into the \textit{R}$\bar{3}$\textit{m} phase after a simple octahedral layer dislocation. Then the intermediate phase \textit{C}${2}$/\textit{m} and the regular \textit{P}${2}$/\textit{m} [Fig. \ref{fig4}(b)] are obtained after the symmetry-breaking by the symmetry-operated transformation. Finally, a simple lattice perturbation is required to induce the transformation of the well-aligned \textit{P}${2}$/\textit{m} phase into the actual \textit{P}${2}$/\textit{m} structure we calculated [Fig. \ref{fig4}(c)].
\begin{equation}
\begin{aligned}
QRP&=
\begin{pmatrix}
 0  & 1/2 & 1  \\
 1  & -1/2 & 0  \\
 0  & 3/4 & 0
\end{pmatrix}
\begin{pmatrix}
 1  & -1 & 0  \\
 0  & -1 & 0  \\
 0  & 0 & -1
\end{pmatrix}
\begin{pmatrix}
 0  & 1 & 2/3  \\
 0  & 0 & 4/3  \\
 1  & 0 & -2/3
\end{pmatrix} \\
&= \begin{pmatrix}
 -1  & 0 & 0  \\
 0  & 1 & 0  \\
 0  & 0 & -1
\end{pmatrix}
= C_{2y}(\textit{P}{2}/\textit{m})
\end{aligned}
\end{equation}
\subsection{\label{sub3-2}Performance Change after Structural Phase Transition under Pressure}

\subsubsection{\label{sub3-2-1}Regulation of crystal anisotropy}
Based on the elastic constants analyzed in Appendix \ref{subB-3-1}, we can get the following results.
First, we predict that the hexagonal phase Baln$_{2}$As$_{2}$ is more compressible in the $ab$-plane, and the octahedral layer in Fig. \ref{fig1}(a) is more susceptible to phase transitions in the $ab$ plane. In contrast, the structural phase transitions of CaIn$_{2}$As$_{2}$ and SrIn$_{2}$As$_{2}$ are in the $c$ direction. This difference explains that experimentally BaIn$_{2}$As$_{2}$ has different structural phases from CaIn$_{2}$As$_{2}$ or SrIn$_{2}$As$_{2}$. The monoclinic phase of $A$In$_{2}$As$_{2}$ has a structural phase transition in the $a$ direction, which is manifested by weaker bonding in the $a$ axis and relatively easy stripping in that direction. Immediately after that, we find that the bulk modulus $B$, shear modulus $G$, and Young's modulus $E$ of the two-phase structures will be effectively regulated by pressure and show different trends (see Table \ref{tabB5}). 

According to the Appendix \ref{subB-3-2}, we have described and analyzed the significance of the various moduli of elasticity and the trend of their evolution under pressure.
The three-dimensional (3D) figures of various elastic moduli ($G$, $E$, linear compression $LC$) in Supplemental Material show that the anisotropic properties of the different structural phases of $A$In$_{2}$As$_{2}$ differ significantly. Various elastic moduli of hexagonal phase (\textit{P}6$_{3}$/\textit{mmc}) $A$In$_{2}$As$_{2}$ under no pressure tend to be crystal isotropic, especially for BaIn$_{2}$As$_{2}$ (see Fig. S1 within the SM\cite{SI}). As shown in the first two rows of Fig. S2 within the SM\cite{SI}, CaIn$_{2}$As$_{2}$ and SrIn$_{2}$As$_{2}$ remain isotropic in their elastic moduli due to too little pressure. However, at 10.555 GPa, the $G$, $E$, and $v$ of the BaIn$_{2}$As$_{2}$ system shift to anisotropy, and the $LC$ tends to remain isotropic (see the third row of Fig. S2 within the SM\cite{SI}). In sharp contrast to the hexagonal phase, the monoclinic (\textit{P}${2}$/\textit{m}) $A$In$_{2}$As$_{2}$ systems exhibit significant crystal anisotropy under no pressure (see Fig. S3-S5 within the SM\cite{SI}). Moreover, the pressure will further enhance the anisotropy of the individual elastic moduli of the monoclinic phase $A$In$_{2}$As$_{2}$ system. 2D projections of the pressure-regulated $G$, $E$, $LC$, and $v$ associated with CaIn$_{2}$As$_{2}$, SrIn$_{2}$As$_{2}$ and BaIn$_{2}$As$_{2}$ are presented in the Supporting Material as Figs. S6-S10\cite{SI}. For a detailed analysis of the anisotropy of these mechanical parameters projected in the $xy$, $yz$, and $xz$ directions, see Appendix \ref{subB-3-3}.

\begin{figure}[!htbp]
	\begin{centering}
		\includegraphics[width=0.5\textwidth]{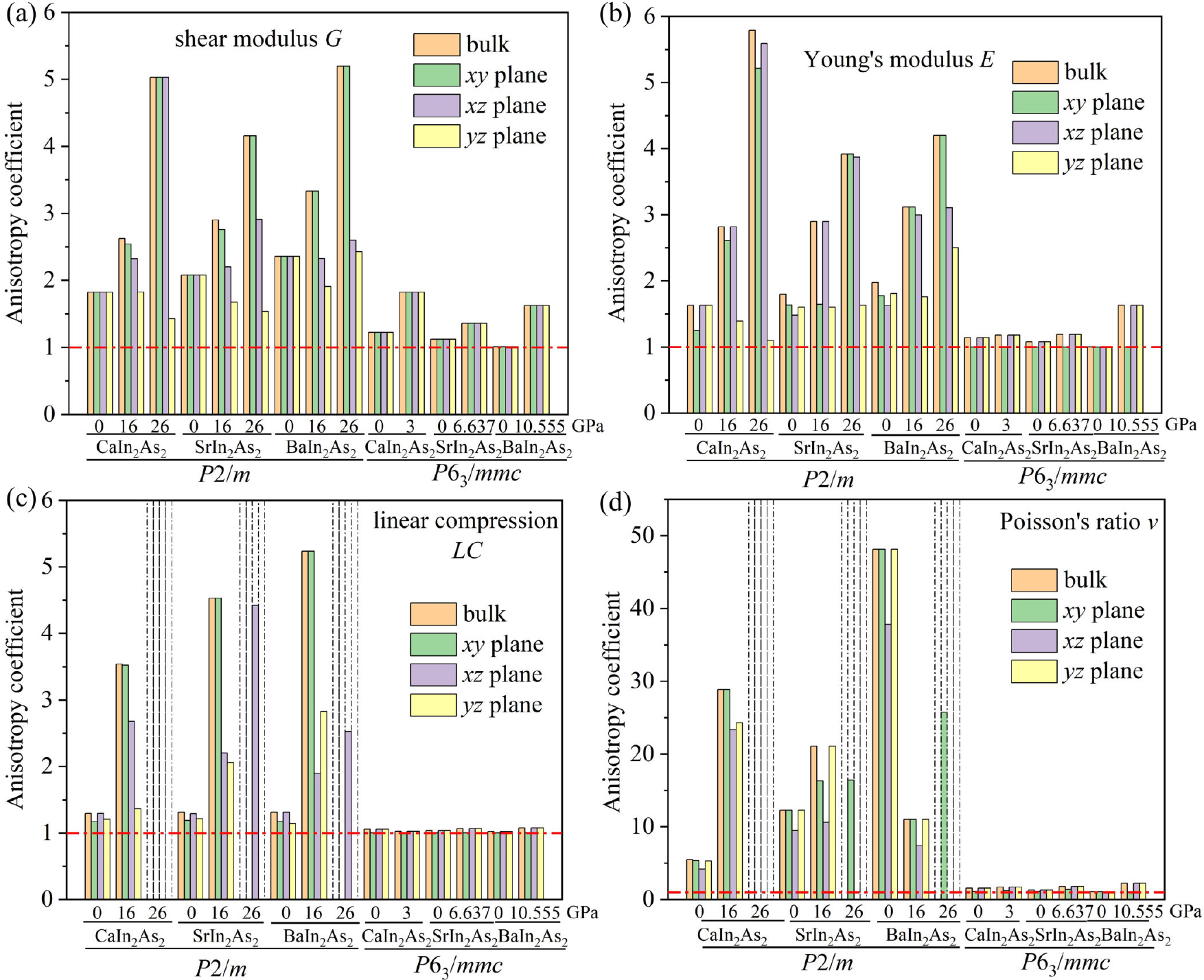}
		\par\end{centering}
	\centering{}\caption{ Bulk anisotropy coefficients and plane anisotropy coefficients for hexagonal phase and monoclinic relative to (a) $G$, (b) $E$, (c) $LC$, (d) $v$ under pressure. The red dashed line indicates the anisotropy coefficient $A$ = 1, representing the complete isotropy. The dashed column indicates infinity anisotropy coefficients, which means that the minimum value of the modulus of elasticity is 0 or negative.}
	\label{fig5}
\end{figure}

We further compared the bulk anisotropy and plane anisotropy coefficients for each elastic modulus of $A$In$_{2}$As$_{2}$ under pressure (see Fig. \ref{fig5}). The anisotropy coefficients of the hexagonal phase (\textit{P}6$_{3}$/\textit{mmc}) mostly exhibit isotropic features and are distributed around the red dashed line in Fig. \ref{fig5}. For $G$ and $E$ of the hexagonal phase, the pressure will somehow enhance their degree of anisotropy. In contrast, the degree of anisotropy of $LC$ and $v$ shows robustness to the pressure. $G$ in hexagonal phase $A$In$_{2}$As$_{2}$ at all pressures and monoclinic phase $A$In$_{2}$As$_{2}$ at 0 GPa exhibit equal bulk and plane anisotropy coefficients [see Fig. \ref{fig5}(a)]. The pressure will break the equilibrium of equal bulk and plane anisotropy coefficients for the monoclinic phase, and the enhancement of the bulk anisotropy mainly comes from the two-plane anisotropy enhancement of $xy$ and $xz$.
In contrast, the $G$ anisotropy of the $yz$ plane is robust for pressure, which does not become significantly more extensive due to pressure, as shown by the yellow rectangle of the \textit{P}${2}$/\textit{m} phase in Fig. \ref{fig5}(a). Similarly, the $yz$-plane anisotropy of Young's modulus $E$ and linear compression $LC$ in the monoclinic phase $A$In$_{2}$As$_{2}$ do not become much larger under pressure modulation. In contrast, the $xy$- or $xz$-plane $E$ and $LC$ anisotropies are an essential reason for the significant increase in the anisotropy of the bulk $E$ and bulk $LC$ [see Figs. \ref{fig5}(b) and \ref{fig5}(c)]. Of interest is the monoclinic phase system where $LC$ and Poisson's ratio $v$ appears to have a minimum value of 0 or even negative at 26 GPa, resulting in an anisotropy of infinity [see the dashed hollow rectangles in Figs. \ref{fig5}(c) and \ref{fig5}(d)]. Although the bulk anisotropy coefficients tend to infinity, there are finite anisotropy coefficients (non-infinity) for SrIn$_{2}$As$_{2}$ (BaIn$_{2}$As$_{2}$) for $LC$ and $v$ in the $xz$-plane and $xy$-plane, respectively.
Moreover, they both undergo a dramatic change in anisotropy under pressure modulation, with the same pattern as the 2D analysis above. Their anisotropy can be described by the two anisotropy constants $A_{U}$ and $A_{L}$ in Table \ref{tabB6}, which can be calculated by Eqs. (\ref{equ15}) and (\ref{equ16}). The values of $A_{U}$ and $A_{L}$ illustrate that BaIn$_{2}$As$_{2}$ in the hexagonal phase is completely isotropic at 0 GPa and that the pressure can substantially enhance the system anisotropy. In addition, the monoclinic phase's anisotropy is stronger than the hexagonal phase's. Calculating the anisotropy constants leads to an assertion consistent with the previous discussion.

\subsubsection{\label{sub3-2-2}Realization of negative Poisson's ratio material}

\begin{figure*}[!htbp]
	\begin{centering}
		\includegraphics[width=0.9\textwidth]{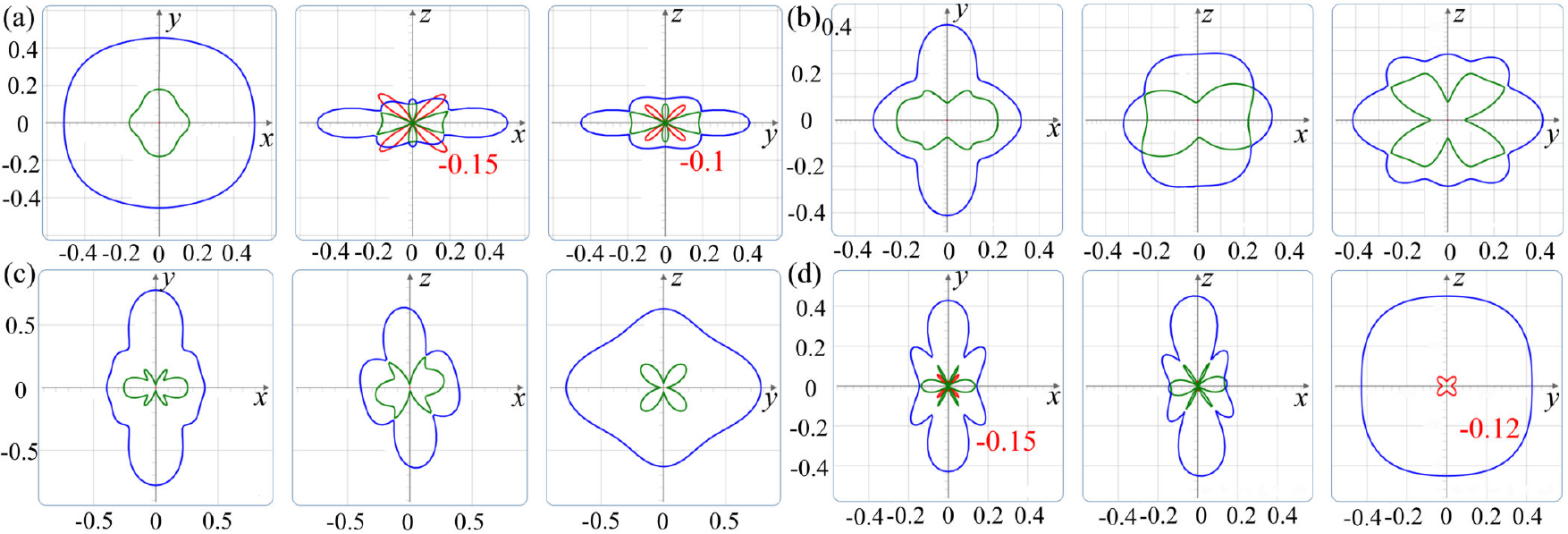}
		\par\end{centering}
	\centering{}\caption{ Poisson's ratio ($v$) of CaIn$_{2}$As$_{2}$ in $xy$, $xz$, and $yz$ planes. (a)-(d) monoclinic phase under $-$ 4 GPa, 0 GPa, 16 GPa, and 26 GPa pressures.}
	\label{fig6}
\end{figure*}

Poisson's ratio is the opposite of transverse strain to axial strain when a material is tensile or compressive in a particular direction. NPR materials, also known as auxetic materials, have several excellent properties because of their unique mechanical structure, including superior fracture resistance, shear resistance, sound and energy absorption, dent resistance, and surface isotropy \cite{Strk2012EffectiveMP,GRUJICIC2013113,https://doi.org/10.1002/pssb.201384250}. Although NPR is allowed by thermodynamics, this property is rare in crystalline solids \cite{1998Natur.392.362B}.
NPR is mainly studied in 2D materials and structures, and it is crucial to design a 3D multilevel system that can exhibit NPR under deformation\cite{NPR}. It is difficult to find materials that can show a negative Poisson ratio under both pressure and tension, and it is even rarer to find materials or structures that can have the same NPR performance under tension and compression stresses\cite{NPR}. 

Using pressure modulation, we observe a NPR phenomenon in $A$In$_{2}$As$_{2}$ with low symmetry \textit{P}${2}$/\textit{m} phase. In the case of CaIn$_{2}$As$_{2}$, for example, the system exhibits a generally NPR behavior in the absence of pressure or at low compressive stresses [see Figs. \ref{fig6}(b) and \ref{fig6}(c)]. At both tensile pressure of $-$4 GPa [see Fig. \ref{fig6}(a)] and compressive pressure of 26 GPa [see Fig. \ref{fig6}(d)], the material rarely exhibits NPR property. We predict that this NPR material can be widely used for many applications in medical devices, cushioning and protective equipment, intelligent sensors, and defence industries.

\subsubsection{\label{sub3-2-3}Thermal Properties Analysis and Enhance in Phase Transition Temperature}

Experimentally synthesized BaIn$_{2}$As$_{2}$ (\textit{P}2/\textit{m}) has different structural phases from CaIn$_{2}$As$_{2}$ and SrIn$_{2}$As$_{2}$ (\textit{P}6$_{3}$/\textit{mmc}). The above first-principles calculations based on absolute zero (T= 0 K) conditions give detailed results of the structural phase transition. However, the thermodynamic physical picture of the structural phases at high temperatures is still blurred. Here, we calculate the dependence of thermodynamic parameters on temperature between the hexagonal phase (\textit{P}6$_{3}$/\textit{mmc}) and the low-symmetry monoclinic phase (\textit{P}2/\textit{m}).
The specific heat at constant volume $C_{\mathbf{v}}$, the vibrational entropy $S_{\mathrm{vib}}(T)$, the internal energy $U_{\mathrm{vib}}(T)$, and the Helmholtz free energy $F(T)$ of individual harmonic oscillator and its difference $\Delta F_{\textit{P}2/\textit{m}-\textit{P}6_{3}/\textit{mmc}}(T)$ between two phases are given as Eqs. (\ref{equ1})-(\ref{equ5}).

To investigate the mechanism of the response of the above-mentioned thermal parameters to temperature under pressure, we compared the thermodynamic curves of the two phases $\textit{P}6_{3}/\textit{mmc}$ and $\textit{P}2/\textit{m}$ under pressure, as shown in Figs. S13 and S14\cite{SI}. As the pressure increases, both phases show an increase in free energy $F$ (red curve), a decrease in entropy $S$ (blue curve), and a convergence of the heat capacity $C_{V}$ to a constant (green curve). As shown by the arrows in the enlarged diagram in the right column of Figs. S13 and S14\cite{SI}, the intersection of heat capacity and entropy tends to move towards higher temperatures as the pressure increases, except for BaIn$_{2}$As$_{2}$ in the \textit{P}2/\textit{m} phase. This exception may be due to lattice distortion inducing a large phonon dispersion spectrum of imaginary frequencies at $\Gamma$ (shown in Fig. S12(i) within the SM\cite{SI}). The phonon frequencies of each system of the \textit{P}2/\textit{m} space group corresponding to Fig. S12 within the SM\cite{SI} at the point $\Gamma$ are shown in Fig. S20(a) within the SM\cite{SI}. It is easy to find that CaIn$_{2}$As$_{2}$ at 0 GPa, BaIn$_{2}$As$_{2}$ at 16 GPa, and $A$In$_{2}$As$_{2}$ at 26 GPa all have large imaginary frequencies. The acoustic and optical branches for each frequency correspond to the irreducible representation and activity (Raman or IR) are compared in Table S1. Unlike other systems where the acoustic branch consists of $A_{u}+2B_{u}$, the acoustic branch of CaIn$_{2}$As$_{2}$ has $B_{g}$ involved in the absence of pressure, and the $B_{u}$ IR activity is squeezed to the fourth branch (-0.95 $cm^{-1}$), leading to the dynamic instability of the system. With the application of pressure, the phonon dispersion spectrum expands and shifts toward high frequencies while CaIn$_{2}$As$_{2}$ opens a gap near 100 $cm^{-1}$.

\begin{figure*}[!htbp]
	\begin{centering}
		\includegraphics[width=0.9\textwidth]{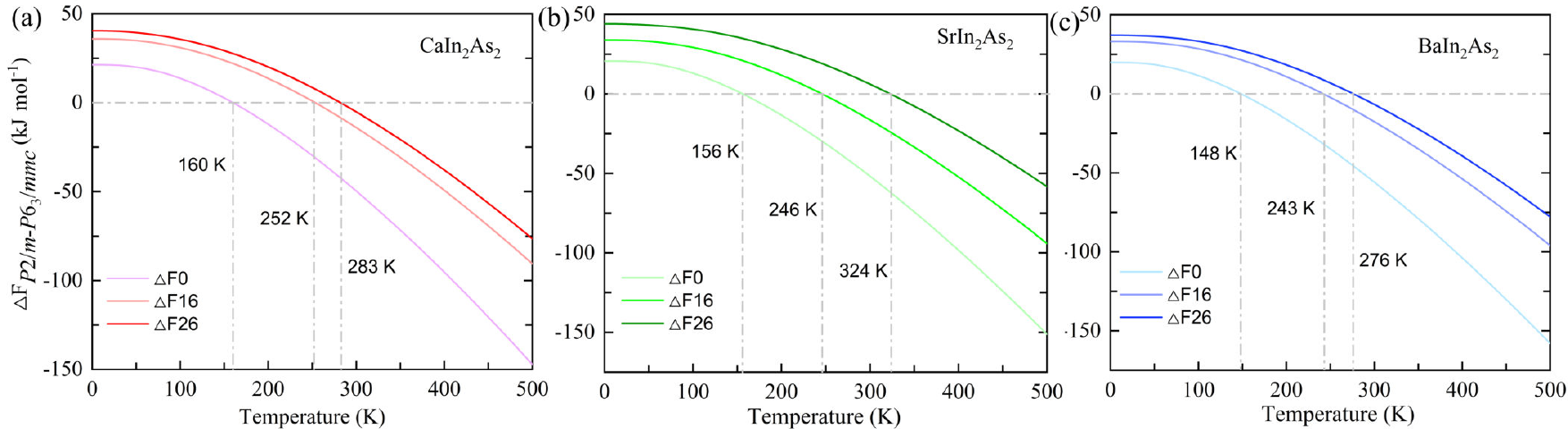}
		\par\end{centering}
	\centering{}\caption{Enlarged diagrams of the Helmholtz free energy difference for $A$In$_{2}$As$_{2}$ monoclinic and hexagonal phases under different pressures.}
	\label{fig7}
\end{figure*}

As shown in Fig. S15(a) within the SM\cite{SI}, the curve of entropy increase indicates that the vibrational entropy favors a monoclinic phase of $A$In$_{2}$As$_{2}$ over a hexagonal phase. The vibrational entropy difference ($\Delta S_{\textit{P}2/\textit{m}-\textit{P}6_{3}/\textit{mmc}}$) between the two phases increases rapidly at low temperatures ($\le$250 K), and then the trend moderates as temperature increases to 3000 K. To quantitatively analyze the vibrational entropy, we give the temperature-dependent characteristic curves of free energy difference including the vibrational entropy [shown in Fig. S15(b) within the SM\cite{SI}]. At low temperatures, the free energy difference between the monoclinic phase and the hexagonal phase is positive, implying that the hexagonal phase is relatively stable. When the temperature increases, the vibrational entropy prefers to stabilize the monoclinic phase, which ($-T \Delta S(T)< 0$) becomes large enough to compensate for the 0 K energy difference ($\Delta E > 0$), prompting the free energy difference to become negative ($\Delta F(T) < 0$) and the phase transition from the hexagonal phase to the monoclinic phase occurs. The transition temperatures are 160 K, 156 K, and 148 K for CaIn$_{2}$As$_{2}$, SrIn$_{2}$As$_{2}$, and BaIn$_{2}$As$_{2}$, respectively. The hexagonal and monoclinic phases are low- and high-temperature phases, respectively, which is inconsistent with the experimentally reported high temperature where CaIn$_{2}$As$_{2}$ and SrIn$_{2}$As$_{2}$ are hexagonal and BaIn$_{2}$As$_{2}$ are monoclinic phases\cite{newref19}. This discrepancy may be caused by defects or lattice distortions under high temperature. 
As shown in Fig. \ref{fig7}, pressure can effectively raise the structural phase transition temperature of $A$In$_{2}$As$_{2}$ beyond absolute zero (273 K). The phase transition temperature decreases with the increased ionicity of the $A$ atoms at 0 and 16 GPa, which is related to the strength of the interatomic chemical bonds. In particular, the SrIn$_{2}$As$_{2}$ system will reach a higher temperature of 324 K at 26 GPa. Our results will provide critical pressure and temperature options for the experimental synthesis of $A$In$_{2}$As$_{2}$ in specific structural phases.

\begin{figure}[!htbp]
	\begin{centering}
		\includegraphics[width=0.5\textwidth]{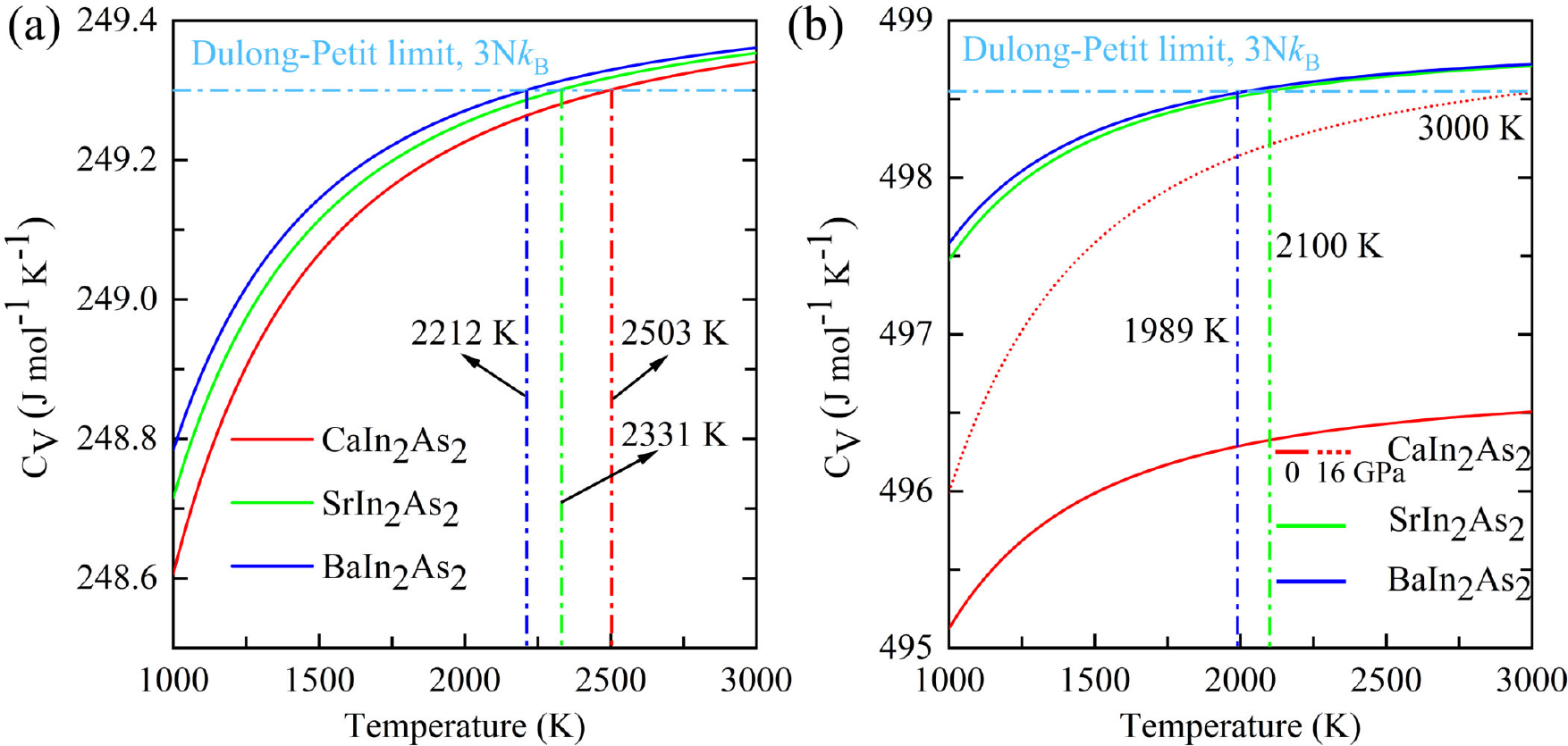}
		\par\end{centering}
	\centering{}\caption{(a) Hexagonal phase and (b) monoclinic phase $C_{V}-T$ curves. The cyan color dashed line indicates the Dulong-Petit limit.}
	\label{fig8}
\end{figure}

Although the heat capacity varies at low temperatures due to pressure (see green curves in Figs. S13 and S14\cite{SI}), the heat capacity of the same phase eventually converges to the same constant independent of pressure and $A$ elements, satisfying the Dulong-petit limit at high temperatures. To observe the change of heat capacity at high temperatures more clearly, we found that the heat capacity of all systems near 1000 K did not reach Dulong-petit limit (see Fig. S17 within the SM\cite{SI}). Except for CaIn$_{2}$As$_{2}$ (\textit{P}2/\textit{m}) under no pressure due to the existence of phonon dispersion at imaginary frequencies causing the heat capacity curve to fall below the 16 GPa case, all of them showed the phenomenon of lowering the high temperature heat capacity worth after pressurization. When the temperature increases to 3000 K, CaIn$_{2}$As$_{2}$ with \textit{P}2/\textit{m} phase at 0 GPa still cannot reach Dulong-petit limit and has the situation of leveling off (see Fig. \ref{fig8}). The rest system of the absence of pressure can cross Dulong-petit limit. However, as shown by the red dashed line in Fig. \ref{fig8}(b), applying a 16 GPa pressure can push the heat capacity curve beyond the Dulong-petit limit.

\subsubsection{\label{sub3-2-4}Zero-group velocity behavior of phonon mode softening and Phonon thermal conductivity prediction}
\begin{figure*}[!htbp]
	\begin{centering}
		\includegraphics[width=0.9\textwidth]{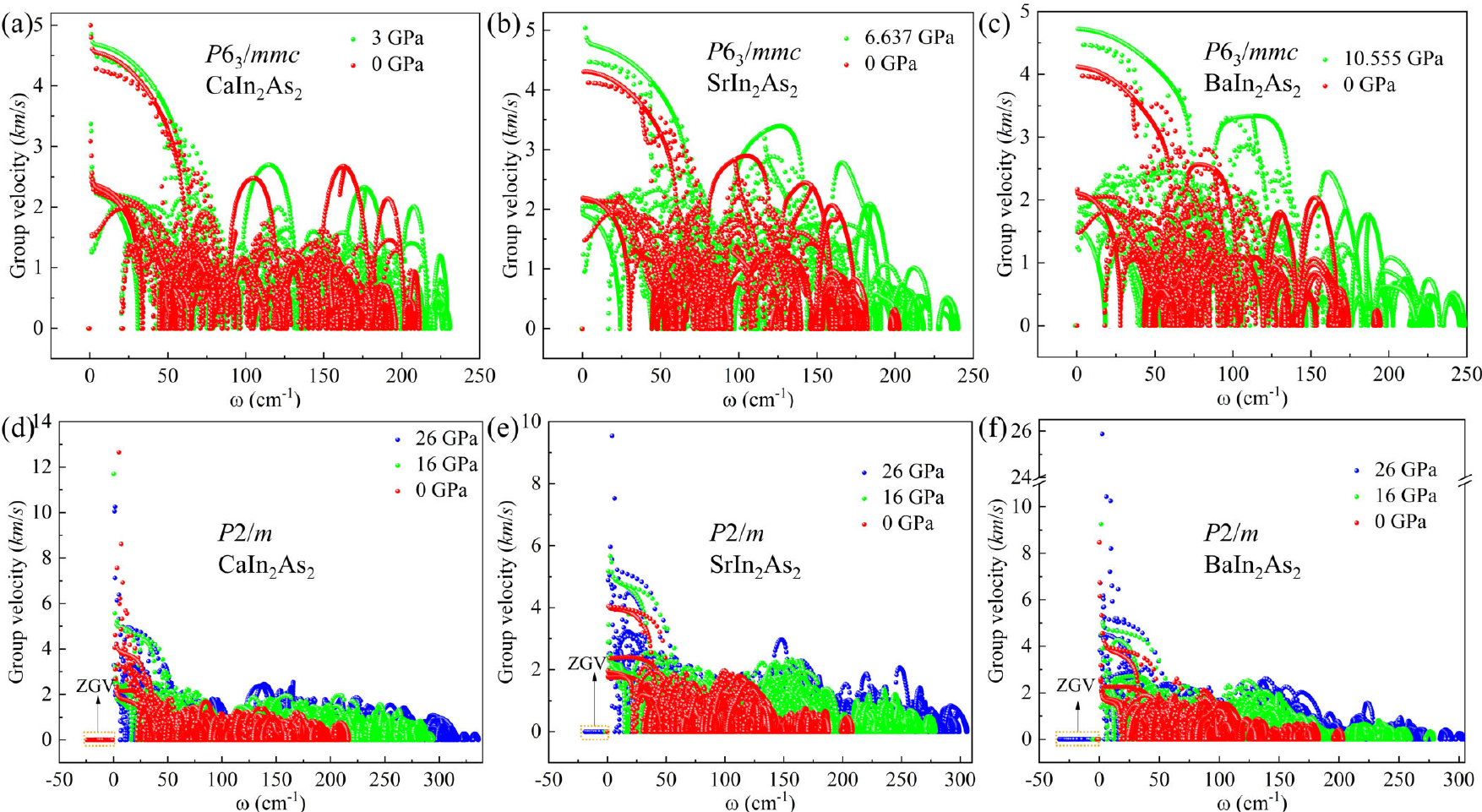}
		\par\end{centering}
	\centering{}\caption{ Group velocity distributions dependent on phonon frequencies for the $\textit{P}6_{3}/\textit{mmc}$ phase (a) CaIn$_{2}$As$_{2}$, (b) SrIn$_{2}$As$_{2}$, (c) BaIn$_{2}$As$_{2}$, and the \textit{P}2/\textit{m} phase (d) CaIn$_{2}$As$_{2}$, (e) SrIn$_{2}$As$_{2}$, and (f) BaIn$_{2}$As$_{2}$ under different pressure.}
	\label{fig9}
\end{figure*}

In order to study in depth the thermal conductivity properties and the sources of thermodynamic instability of the $A$In$_{2}$As$_{2}$ material at high pressures, we calculated phonon group velocities, as shown in Fig. \ref{fig9}. Figures \ref{fig9}(a)-\ref{fig9}(c) demonstrate that the hexagonal phase ($\textit{P}6_{3}/\textit{mmc}$ space group) is both stable in the absence of pressure and at the induced zero bandgap pressure. The pressure induces the group velocity towards lower and higher frequencies, behaving more divergent. In addition, both two phases of $A$In$_{2}$As$_{2}$ show a tendency for the group velocity to become larger in the medium and high-frequency regions with increasing pressure (see Fig. \ref{fig9}). Moreover, the low-frequency acoustic branch mainly contributes to the phonon thermal conductivity of all systems. However, compared to the hexagonal phase, the monoclinic phase of $A$In$_{2}$As$_{2}$ has larger group velocities in the low-frequency region, and the group velocities in the medium and high-frequency areas are all roughly distributed in the range of 2-3 $km/s$. The pressure drives a virtual frequency in the low-frequency region because of the appearance of softened phonon modes, which induces a ZGV [see Figs. \ref{fig9}(d)-\ref{fig9}(f)].

The phonon thermal conductivity depends on the group velocity with the relation $\kappa = Cv_{g}^2\tau$ , where $\tau$ is the average relaxation time. The thermal conductivity can be initially predicted from v$_{g}^2$. Figures S18 and S19\cite{SI} give images of the frequency dependence of the squared group velocity v$_{gi}^2 (i=x, y, z)$ in the three directions of the hexagonal and monoclinic phases $A$In$_{2}$As$_{2}$ under pressure. Closely related to the crystal structure, the $\textit{P}6_{3}/\textit{mmc}$ phase has similar group velocity evolution curves in the $x$ and $y$ directions. Therefore, for the $\textit{P}6_{3}/\textit{mmc}$ phase, we refer to the thermal conductivity transported within the octahedral inner layer as the in-plane thermal conductivity. Along the $z$-direction, we refer to the out-of-plane thermal conductivity. The out-of-plane thermal conductivity of the hexagonal phase $A$In$_{2}$As$_{2}$ is slightly larger than the in-plane thermal conductivity. In the low-frequency region, thermal conduction is more favoured along out-of-plane. However, the fluctuation phenomenon of the out-of-plane thermal conductivity is more pronounced, with zero thermal conductivity behavior in specific frequency regions, and thermal conductivity is frequency selective. Therefore, due to the octahedral lattice's hindrance, the out-of-plane thermal conduction behavior of the hexagonal phase $A$In$_{2}$As$_{2}$ is not as easy. For the monoclinic phase, as shown in Fig. S19 within the SM\cite{SI} $A$In$_{2}$As$_{2}$ has thermal conduction anisotropy in three directions, and the $y$ direction is the main direction of thermal conduction (for CaIn$_{2}$As$_{2}$). This is because both the \textit{P}2/\textit{m} phase structure along the $x$ and $z$ directions must traverse the $A$-As octahedral lattice [see the lattice structure in Fig. \ref{fig1}(c)]. In contrast, along the $y$ direction, thermal conductivity is possible through the interstices of the octahedral lattice. For the monoclinic phase, the contribution of thermal conduction in the $z$ direction is also significant to a certain extent. In summary, the $x$-direction is the most difficult direction for thermal conduction in the monoclinic phase, which is related to the smallest C$_{11}$ elastic constant (see Fig. \ref{figB2})

The pressures all enhance the group velocity for $A$In$_{2}$As$_{2}$ systems, leading to a shift of the group velocity towards high and low frequencies. The ZGV phenomenon resulting from shifting the monoclinic phase phonon spectrum towards lower frequencies under pressure directly reflects the softening of the phonon modes. The phonon frequency distribution of the monoclinic phase under pressure and the structure with atomic sites are given as shown in Fig. S20 within the SM\cite{SI}. We focus on the CaIn$_{2}$As$_{2}$ (0, 26 GPa), SrIn$_{2}$As$_{2}$ (26 GPa), and BaIn$_{2}$As$_{2}$ (26 GPa) systems that produce significant imaginary frequencies. The group velocity at the imaginary frequency (IFGV) of CaIn$_{2}$As$_{2}$ under pressure absences is mainly contributed by the In-$2n$ position (red dashed circle in Fig. S20(b) within the SM\cite{SI}) and the As-$2m$ position (blue dashed circle in Fig. S20(b) within the SM\cite{SI}). Most of the IFGV of CaIn$_{2}$As$_{2}$ under 26 GPa originates from not only the atomic contributions from the two  Wyckoff sites mentioned above, but also the In-$2n$ site of the cyan colour and the green-coloured As-$2n$ site. For the IFGV of SrIn$_{2}$As$_{2}$ and BaIn$_{2}$As$_{2}$ at 26 GPa, there is also a contribution from the $A$ atom in addition to the In and As atom contributions. The IFGV of SrIn$_{2}$As$_{2}$ under 26 GPa is mainly contributed by the Cyan-coloured In-$2n$ site, the blue-coloured As-$2m$ site and the $A$$_{2}$-$1c$ site. BaIn$_{2}$As$_{2}$, on the other hand, is primarily contributed by the black-coloured In-$2m$ site, the green-coloured As-$2n$ site, the $A$$_{1}$-$1d$ site and the $A$$_{2}$-$1c$ site. The main contributing atoms to the ZGV phenomenon produced by the softening of the phonon vibrational modes are also presented in Fig. S20(c) within the SM\cite{SI}. We can clearly find that the phonon mode softening is critically due to atomic vibrations across the zero frequency near the In-In chain ( [101] direction in Fig. S20 within the SM\cite{SI}). As seen in Table S1\cite{SI}, for the CaIn$_{2}$As$_{2}$ (0 GPa) system, the relatively large imaginary frequencies (below -0.08 cm$^{-1}$) are mainly contributed by IR-active A$_{u}$ and Raman-active B$_{g}$. For $A$In$_{2}$As$_{2}$ under 26 GPa, on the other hand, the virtual frequencies are all mainly contributed by B$_{u}$. Therefore, softening the phonon modes at high pressure weakens the IR vibrational modes, A$_{u}$ and B$_{u}$. It is interesting to note that the strange imaginary frequency of CaIn$_{2}$As$_{2}$ at 0 GPa also originates from the appearance of the Raman vibrational mode B$_{g}$, which should not have appeared in the acoustic branch.
We can predict that along between the $A$-As octahedral layers (In-In atomic gaps) is the direction of maximum probability of phonon softening and thermal conduction in the monoclinic phase $A$In$_{2}$As$_{2}$.

\section{\label{sec4}Conclusion}
Based on DFT calculations, we have predicted the structural phase transition of $A$In$_{2}$As$_{2}$ materials under pressure and characterised their mechanical and thermal properties. Firstly, enthalpy of formation and phonon spectroscopy calculations confirm the structural phase transition of $A$In$_{2}$As$_{2}$ under pressure. Moreover, the low-pressure phase of Ca(Sr)In$_{2}$As$_{2}$ materials is hexagonal, while the high-pressure phase is monoclinic. But BaIn$_{2}$As$_{2}$ always prefers to form monoclinic phases. Next, we deeply analyse the symmetries of different space groups, propose the structural phase transition path of $\textit{P}6_{3}/\textit{mmc}$$\rightarrow$\textit{R}$\bar{3}$\textit{m}$\rightarrow$\textit{C}2/\textit{m}$\rightarrow$\textit{P}2/\textit{m} with \textit{C}2/\textit{m} as the intermediate phase, and establish the physical correlation behind the structural phase transition. 

We then also obtain a variety of elastic moduli based on the elastic stiffness matrix and further analyse the crystal anisotropy, chemical bonding properties, hardness, toughness and other mechanical properties of the $\textit{P}6_{3}/\textit{mmc}$ phase and the \textit{P}2/\textit{m} phase $A$In$_{2}$As$_{2}$. Among them, we have deeply investigated the crystal anisotropy transition of $A$In$_{2}$As$_{2}$ series materials based on pressure. Pressure will induce a transition from isotropy to anisotropy in the $A$In$_{2}$As$_{2}$ of the hexagonal phase. Pressure will also enhance the crystal anisotropy in the monoclinic phase. In addition, the bulk anisotropy of these mechanical parameters ($G$, $E$, $LC$, $v$) depends differently on the plane anisotropy of the $xy$,$yz$, and $xz$ planes. We also find that pressure will induce a transition from brittle to ductile in the $A$In$_{2}$As$_{2}$ of the monoclinic and hexagonal phases. And it is found that $A$In$_{2}$As$_{2}$ can be transformed into NPR materials under both compressive and tensile stresses. At the same time, we predict the hardness of different structural phases of $A$In$_{2}$As$_{2}$ that depend on the band gap. 

On the other hand, we postulate that downward pressure can effectively raise materials' structural phase transition temperature and report their thermal properties such as heat capacity, entropy and free energy. Pressure is favoured to enhance the heat capacity profile of the softened monoclinic CaIn$_{2}$As$_{2}$ to reach the Dulong-Petit limit. Thus, we determined that Ca(Sr)In$_{2}$As$_{2}$ is hexagonal at low pressure. BaIn$_{2}$As$_{2}$ enjoys a monoclinic phase but will be in the NPMP phase with similar energies of the monoclinic and hexagonal phases if stretched. At low temperatures, $A$In$_{2}$As$_{2}$ materials prefer to form the hexagonal phase, but they will transform into a monoclinic phase under high temperatures. Moreover, the pressure is favorable to increase the transition temperature of the structural phase. A theoretical basis is laid for a better study of the thermoelectric properties of $A$In$_{2}$As$_{2}$. In a nutshell, our study confirms the mechanical properties and thermal behavior behind the structural phase transition of this family of materials.

\section{Acknowledgments}
We acknowledge the financial support by the National Natural Science Foundation of China (No. 11874113) and the Natural Science Foundation of Fujian Province of China (No. 2020J02018).
The work was carried out at National Supercomputer Center in Tianjin,and the calculations were performed on TianHe-1(A).

\appendix
\counterwithin{figure}{section}
\counterwithin{table}{section} 
\counterwithin{equation}{section} 

\begin{appendices}

\section{\label{secA}COMPUTATIONAL DETAILS}

\subsection{\label{subA-1}Phonon and Thermodynamic Properties Calculation}
For the phonon calculation, the density functional perturbation theory (DFPT) in PHONOPY\cite{TOGO20151} was applied to combine with VASP in the structures of the \textit{P}6$_{3}$/\textit{mmc}, \textit{R}$\bar{3}$\textit{m}, and \textit{P}2/\textit{m} space groups by the 2$\times$2$\times$1, 2$\times$2$\times$2, and 1$\times$2$\times$1 supercells, respectively. 

Thermodynamic properties, including heat capacity, internal energy, entropy, and Helmholtz free energy, were calculated by the following equations\cite{PhysRevMaterials.3.033601,PhysRevB.88.024119}:
\begin{equation}
\begin{aligned}
C_{\mathbf{v}} & =\frac{k_{B}}{N_{q}} \sum_{\boldsymbol{q}, j}\left(\frac{\hbar \omega_{q j}}{2 k_{B} T}\right)^{2} \operatorname{cosech}^{2}\left(\frac{\hbar \omega_{q j}}{2 k_{B} T}\right),
\end{aligned}
\label{equ1}
\end{equation}

\begin{equation}
\begin{aligned}
U_{\mathrm{vib}}(T) & =\frac{1}{N_{q}} \sum_{q, j} \hbar \omega_{q j}\left[\frac{1}{e^{\hbar \omega_{q j} / k_{B} T}-1}+\frac{1}{2}\right]
\end{aligned}
\label{equ2}
\end{equation}

\begin{equation}
\begin{aligned}
S_{\mathrm{vib}}(T) & =\frac{k_{B}}{N_{q}} \sum_{q, j}\left[\frac{}{} \frac{\hbar \omega_{q j}}{k_{B} T(e^{\frac{\hbar \omega_{q j}}{k_{B} T}}-1)}-\ln \left(1-e^{\frac{-\hbar \omega_{q j}} {k_{B} T}}\right)\right]
\end{aligned}
\label{equ3}
\end{equation}

\begin{equation}
\begin{aligned}
F(T) & =\frac{1}{N_{q}} \sum_{q, j}\left[\frac{\hbar \omega_{q j}}{2}+k_{B} T \ln \left(1-e^{-\hbar \omega_{q j} / k_{B} T}\right)\right]
\end{aligned}
\label{equ4}
\end{equation}

\begin{equation}
\begin{aligned}
\Delta F_{\textit{P}2/\textit{m}-\textit{P}6_{3}/\textit{mmc}}(T)=\Delta E+\Delta U_{\mathrm{vib}}(T)-T \Delta S(T).
\end{aligned}
\label{equ5}
\end{equation}
where $k_{B}$ is the Boltzmann constant, $N_{q}$ is the number of wave vectors $q$, and $\omega_{q j}$ is the vibrational frequency of the phonon mode $qj$. $\Delta$ in equation (\ref{equ5}) denotes each physical parameter difference, where $\Delta$E is the total energy difference calculated by VASP.

\subsection{\label{subA-2}Mechanical Properties Characterization}
\subsubsection{\label{subA-2-1}Elastic Moduli and Mechanical Stability Criteria}
The elastic modulus formulas for the hexagonal and monoclinic phases are from Ref. \cite{doi:10.1021/acsami.2c11350} and Ref. \cite{doi:10.1063/1.327803}, respectively. The mechanical stability criterion is from Ref. \cite{Nye}. The Voigt Reuss-Hill\cite{Hill} approximation is the arithmetic mean of the Voigt\cite{Voigt} and Reuss bounds\cite{Reuss}. B denotes the bulk modulus, G denotes the shear modulus, E denotes Young's modulus, and $v$ denotes Poisson's ratio.
According to the Voigt-Reuss-Hill approximation\cite{Hill}, $X_H = (1/2)(X_R+X_V)$, $X$ = $B$, $G$. Furthermore, Young's modulus E and Poisson's ratio $v$ are derived from Eq. (\ref{equ6}):

\begin{equation}
\begin{aligned}
E = \frac{9BG}{3B + G},
v = \frac{3B-2G}{6B+2G}.
\end{aligned}
\label{equ6}
\end{equation}

The independent elastic stiffness constants $C_{ij}$ of hexagonal phase include $C_{11}$, $C_{33}$, $C_{44}$, $C_{12}$, and $C_{13}$. The modulus can be described as follows:

\begin{equation}
\begin{aligned}
B_{V}=\frac{1}{9}[2(C_{11}+C_{12})+4 C_{13}+C_{33}], 
\end{aligned}
\label{equ7}
\end{equation}

\begin{equation}
\begin{aligned}
G_{V}=\frac{1}{30}(M+12 C_{44}+12 C_{66}),
\end{aligned}
\label{equ8}
\end{equation}

\begin{equation}
\begin{aligned}
B_{R}=\frac{C^{2}}{M}, 
\end{aligned}
\label{equ9}
\end{equation}

\begin{equation}
\begin{aligned}
G_{R}=\frac{\frac{5}{2}(C^{2} C_{44} C_{66})}{3 B_{V} C_{44} C_{66}+C^{2}(C_{44}+C_{66})}, 
\end{aligned}
\label{equ10}
\end{equation}

where
$$
\begin{aligned}
M=C_{11}+C_{12}+2 C_{33}-4 C_{13}, \\
C^{2}=(C_{11}+C_{12}) C_{33}-2 C_{13}^{2}.
\end{aligned}
$$

The mechanical stability criteria are given via
$$
C_{44}>0, \quad C_{11}>\left|C_{12}\right|, \quad\left(C_{11}+2 C_{12}\right) C_{33}>2 C_{13}^2 .
$$

As for monoclinic phase, the independent $C_{ij}$ can be indicated to $C_{11}$, $C_{22}$, $C_{33}$, $C_{44}$, $C_{55}$, $C_{66}$, $C_{12}$, $C_{13}$, $C_{23}$, $C_{15}$, $C_{25}$, $C_{35}$, and $C_{64}$. The modulus can be described as follows:

\begin{equation}
\begin{aligned}
B_V=\frac{1}{9}[C_{11}+C_{22}+C_{33}+2(C_{12}+C_{13}+C_{23})],
\end{aligned}
\label{equ11}
\end{equation}

\begin{equation}
\begin{aligned}
G_V=&\frac{1}{15}[C_{11}+C_{22}+C_{33}+3(C_{44}+C_{55}+C_{66})-\\
&(C_{12}+C_{13}+C_{23})]. 
\end{aligned}
\label{equ12}
\end{equation}

\begin{equation}
\begin{aligned}
B_{R}=&\Omega[a(C_{11}+C_{22}-2 C_{12})+b(2 C_{12}-2 C_{11}-C_{23})\\
&+c(C_{15}-2 C_{25})+d(2 C_{12}+2 C_{23}-C_{13}-2 C_{22})\\
&+2 e(C_{25}-C_{15})+f]^{-1},
\end{aligned}
\label{equ13}
\end{equation}

\begin{equation}
\begin{aligned}
G_{R}=&15\{[4a(C_{11}+C_{22}+C_{12})+b(C_{11}-C_{12}-C_{23})+\\
&c(C_{15}+C_{25})+d(C_{22}-C_{12}-C_{23}-C_{13})+\\
&e(C_{15}-C_{25})+f]/\Omega+3[\frac{g}{\Omega}+\frac{C_{44}+C_{66}}{C_{44} C_{66}-C_{64}^{2}}]\}^{-1},
\end{aligned}
\label{equ14}
\end{equation}

\begin{equation}
\begin{aligned}
a=&C_{33} C_{55}-C_{35}^{2},b=C_{23} C_{55}-C_{25} C_{35},\\
 c=&C_{13} C_{35}-C_{15} C_{33}, d=C_{13} C_{55}-C_{15} C_{35},\\
e=&C_{13} C_{25}-C_{15} C_{23},\\
f= & C_{11}(C_{22} C_{55}-C_{25}^{2})-C_{12}(C_{12} C_{55}-C_{15} C_{25}) \\
& +C_{15}(C_{12} C_{25}-C_{15} C_{22})+C_{25}(C_{23} C_{35}-C_{25} C_{33}), \\
g=&  C_{11} C_{22} C_{33}-C_{11} C_{23}^{2}-C_{22} C_{13}^{2}-C_{33} C_{12}^{2}+2 C_{12} C_{13} C_{23},\\
\Omega=& 2[C_{15} C_{25}(C_{33} C_{12}-C_{13} C_{23})+C_{15} C_{35}(C_{22} C_{13}-\\
&C_{12} C_{23}+C_{25} C_{35}(C_{11} C_{23}-C_{12} C_{13})]-[C_{15}^{2}(C_{22} C_{33}-\\
&C_{23}^{2})+C_{25}^{2}(C_{11} C_{33}-C_{13}^{2})+C_{35}^{2}(C_{11} C_{22}-C_{12}^{2})]+g C_{55} .
\nonumber
\end{aligned}
\end{equation}

The criteria for mechanical stability are given via
$$
\begin{gathered}
C_{11}>0, C_{22}>0, C_{33}>0, C_{44}>0, C_{55}>0, C_{66}>0, \\
 [C_{11}+C_{22}+C_{33}+2(C_{12}+C_{13}+C_{23})]>0, \\
(C_{33} C_{55}-C_{35}^2)>0, (C_{44} C_{66}-C_{46}^2)>0, \\
(C_{22}+C_{33}-2 C_{23})>0,\\
[C_{22}(C_{33} C_{55}-C_{35}^{2})+2 C_{23} C_{25} C_{35}-C_{23}^{2} C_{55}
-C_{25}^{2} C_{33}]>0,\\
\{2 [C_{15} C_{25}(C_{33} C_{12}-C_{13} C_{23})+C_{15} C_{35}(C_{22} C_{13}-C_{12} C_{23}) \\
 +C_{25} C_{35}(C_{11} C_{23}-C_{12} C_{13})]-[C_{15}^{2}(C_{22} C_{33}-C_{23}^{2}) \\
+C_{25}^{2}(C_{11} C_{33}-C_{13}^{2})+C_{35}^{2}(C_{11} C_{22}-C_{12}^{2})]+gC_{55} \}>0 .
\end{gathered}
$$

\subsubsection{\label{subA-2-2}Crystal anisotropy calculation}
 Since Zener anisotropy\cite{Zener} and Chung-Buessem anisotropy\cite{doi:10.1063/1.1709819} indices are only applicable to cubic crystals, we used the universal anisotropy index $A_{U}$\cite{PhysRevLett.101.055504} and log-Euclidean anisotropy index $A_{L}$\cite{doi:10.1063/1.4962996} for the anisotropy analysis of \textit{P}6$_{3}$/\textit{mmc} and \textit{P}2/\textit{m} phases. $A_{U}$ takes into account all the stiffness coefficients to define the anisotropy, exploiting the tensor nature of the elastic stiffness. The specific expression is shown in Eq. (\ref{equ15}). The expression for $A_{L}$ with respect to the modulus of elasticity is given by Eq. (\ref{equ16}).

\begin{equation}
\begin{aligned}
A_{U}=5 \frac{G_{V}}{G_{R}}+\frac{B_{V}}{B_{R}}-6 
\end{aligned}
\label{equ15}
\end{equation}

\begin{equation}
\begin{aligned}
A_{L} =\sqrt{ \left[\ln \left(\frac{B_{V}}{B_{R}}\right)\right]^{2} + 5\left[\ln \left(\frac{G_{V}}{G_{R}}\right)\right]^{2} }
\end{aligned}
\label{equ16}
\end{equation}

The value of the anisotropy parameter ($A_{U}$ and $A_{L}$) is $\ge$ 0. They characterize the strength of the crystal anisotropy, and their convergence to zero implies that crystal isotropy.

\subsubsection{\label{subA-2-3}Bonding information calculation}
The Kleinman parameter ($\xi$) allows evaluation of the stability of the solid under stretching or bending\cite{PhysRev.128.2614}, which is defined as:
$$
\xi=\frac{C_{11}+8C_{12}}{7C_{11}-2C_{12}}
$$
$\xi$ = 0 and 1 imply that bond bending and stretching will be dominated, respectively. 

The Cauchy pressure ($P_{C}$) can also be used to describe the brittleness and ductility of a metal or compound. For hexagonal crystal systems, it is defined as $P_{C}^{a}$=$C_{13}-C_{44}$ and $P_{C}^{b}$=$C_{12}-C_{66}$\cite{BAO2019106027}.

\subsubsection{\label{subA-2-4}Hardness prediction}

First-principles calculations provide a good assessment of the various mechanical properties of a solid. However, DFT does not give a reasonable evaluation of hardness directly. We predict hardness based on the following semi-empirical relationships to describe the mechanical behavior of $A$In$_{2}$As$_{2}$ fullyfully\cite{83JIANG20112287,84JIANG20112287,85Jiang2010,86MIAO20111559,87CHEN20111275},

\begin{equation}
\begin{gathered}
H_{1a}=0.1475 G, H_{1b}=0.0607 E  \qquad \text{\citep{83JIANG20112287}},\\
H_{2}=0.1769 G-2.899  \qquad  \qquad  \qquad \text{\citep{84JIANG20112287}},\\
H_{3}=0.0635 E \qquad  \qquad \qquad \qquad \quad \text{\citep{85Jiang2010}}, \\
H_{4}=\frac{(1-2v) B}{6(1+v)} \qquad  \qquad \qquad \qquad \text{\citep{86MIAO20111559}}, \\
H_{5}=2(\frac{G^3} {B^2})^{0.585}-3 \qquad  \qquad \qquad \text{\citep{87CHEN20111275}}. 
\end{gathered}
\label{equ17}
\end{equation}

In addition, Ivanovskii is well-placed to summarize these semi-experiences\cite{82IVANOVSKII2013179}. Furthermore, Sobhit Singh calculated the hardness of various materials and compared it with experimental data to choose the semi-empirical calculation of the most appropriate hardness based on the material's space group and band gap\cite{singh2021mechelastic}(see Table \ref{tabA1}). 

\begin{table*}
\centering
\caption{Hardness prediction results for different structural phases dependent on the band gap.}
\begin{tabular}{cccccc}
\hline
\hline
{Type of   material} & {General} & {Cubic} & {Hexagonal} & {Orthorhombic} & {Rhombohedral} \\ \hline
Insulator                                & \multirow{2}{*}{$H_{2}$}       & \multirow{2}{*}{$H_{2}$}        & \multirow{2}{*}{$H_{1b}$}           & \multirow{2}{*}{$H_{2}$}               & \multirow{2}{*}{$H_{2}$}               \\
(Eg \textgreater 2 eV)                   &                              &                            &                                &                                   &                                   \\
Semiconductor                            & \multirow{2}{*}{$H_{5}$}          & \multirow{2}{*}{$H_{5}$}        & \multirow{2}{*}{$H_{1b}$,$H_{3}$}        & \multirow{2}{*}{\_}               & \multirow{2}{*}{$H_{2}$}               \\
(0 \textless Eg \textless 2 eV)          &                              &                            &                                &                                   &                                   \\
Metal                                    & \multirow{2}{*}{$H_{4}$}          & \multirow{2}{*}{$H_{1a}$}       & \multirow{2}{*}{$H_{4}$}            & \multirow{2}{*}{H4}               & \multirow{2}{*}{$H_{4}$}               \\
(Eg = 0)                                 &                              &                            &                                &                                   &    \\
\hline
\hline
\end{tabular}
\label{tabA1}
\end{table*}

ELATools\cite{yalameha2021elatools}, MechElastic\cite{singh2018mechelastic,singh2021mechelastic} and ELATE\cite{Gaillac2016} programs were used for the calculation of mechanical parameters and visualization of the modulus.

\section{\label{secB}Additional RESULTS}
\subsection{\label{secB-1}Evolution of Lattice Parameters under Pressure}
\begin{figure*}[!hbtp]
	\begin{centering}
		\includegraphics[width=1\textwidth]{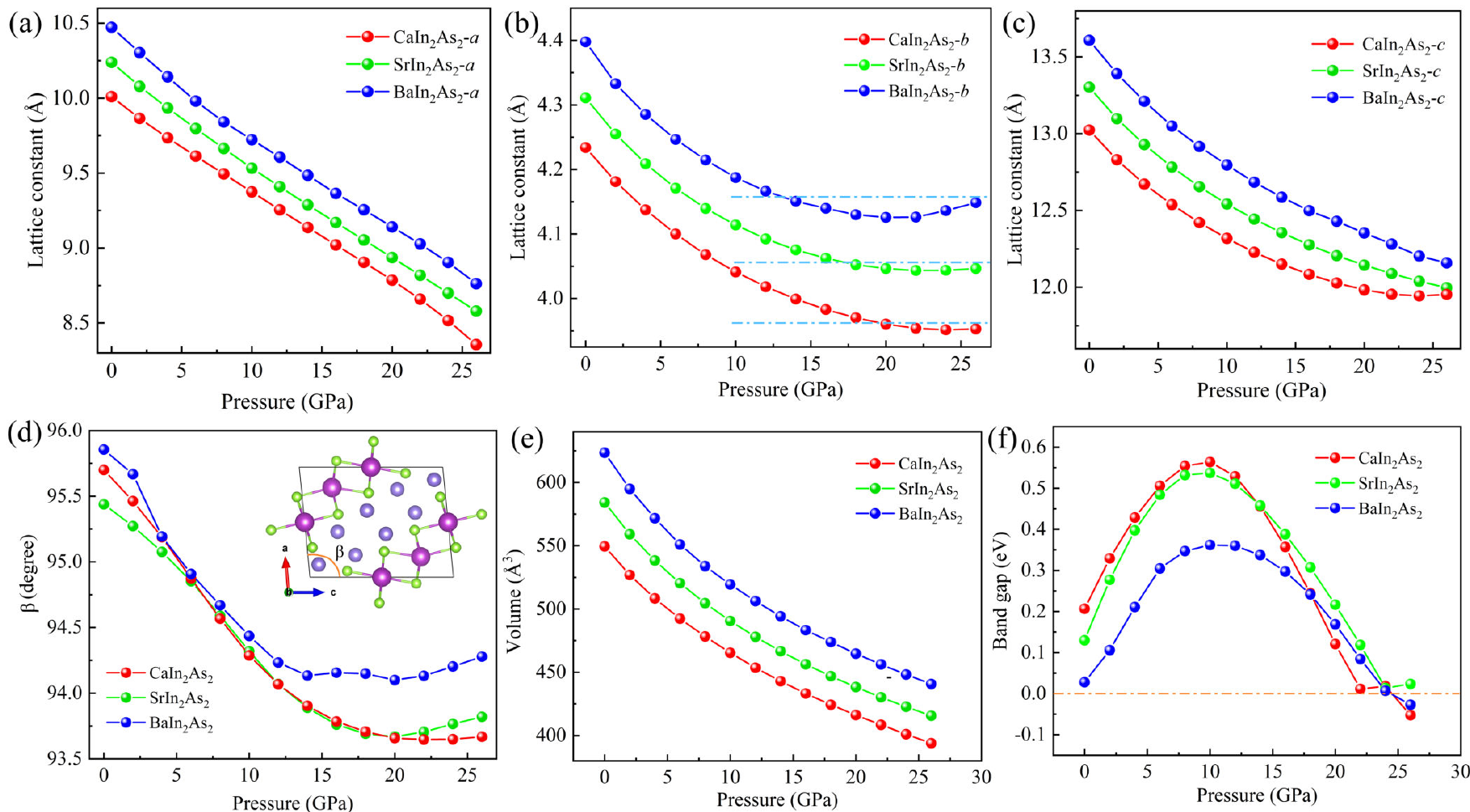}
		\par\end{centering}
	\centering{}\caption{Pressure-dependent evolution curves for $A$In$_{2}$As$_{2}$ of monoclinic phases (\textit{P}2/\textit{m} space group). (a)-(c) Lattice constants, (d) $ac$-plane angle, (e) volume, and (f) band gap.}
	\label{figB1}
\end{figure*}

The application of pressure will first directly change the lattice parameters of the material. The \textit{R}$\bar{3}$\textit{m} and \textit{P}6$_{3}$/\textit{mmc} space groups, which also belong to the hexagonal crystal system, have similar pressure-dependent lattice parameter evolution patterns. As the reported earlier, the lattice constants of $A$In$_{2}$As$_{2}$ materials with the \textit{P}6$_{3}$/\textit{mmc} space group both decrease with increasing pressure. In addition, the bond angles of the hexagonal crystal system are robust to pressure, always maintaining $\alpha = 90\degree$, $\beta = 90\degree$, $\gamma = 120\degree$. However, the bond angle $\beta$ of monoclinic crystal systems is very sensitive to pressure. As shown in Fig. \ref{figB1}(d), the bond angles of the three systems first decrease then increase with increasing pressure, especially for the SrIn$_{2}$As$_{2}$ and BaIn$_{2}$As$_{2}$ systems, where this evolution regular is more obvious.
The lattice constants \textit{a},\textit{b}, and \textit{c} all show a general trend of becoming smaller with the applied positive pressure [see Fig. \ref{figB1}(a)-\ref{figB1}(c)]. The lattice constant \textit{a} is almost linear with pressure, and the lattice constants \textit{b} and \textit{c} are gradually decreasing curves. It is important to note that the lattice constant \textit{b} of BaIn$_{2}$As$_{2}$ becomes larger again under high pressure. As shown in Fig. \ref{figB1}(e), the volume-pressure curve fully reflects the effect of pressure. As with the lattice parameters, the volume shows a positive correlation with the radius of the $A$ atom (BaIn$_{2}$As$_{2}$ is the largest and CaIn$_{2}$As$_{2}$ the smallest).
As shown in Fig. \ref{figB1}(f), the band gaps of all three $A$In$_{2}$As$_{2}$ systems show a trend of increasing and then decreasing under pressure, and all have a band gap maximum around 10 GPa. Narrow band gaps are often accompanied by the non-trivial topological properties of band inversion.

\subsection{\label{secB-2}Symmetry Transformation of Point Group}

For the space group \textit{R}$\bar{3}$\textit{m} with point group D$_{3d}$ ($-3m$):
$$\Gamma_{acoustic} = A_{2u} + E_{u} $$
$$\Gamma_{optic} = 2A_{1g} + 2A_{2u} + 2E_{g} + 2E_{u}$$
In total, there are 15 vibrational modes, 5 nondegenerate $A_{1g} and A_{2u}$ modes, and 5 doubly degenerate $E_{g}$ and $E_{u}$ modes. Among them, optical vibrations $2A_{1g}$ + $2E_{g}$ are Raman (R) active, while optical modes $2A_{2u}$ + $2E_{u}$ are infrared Raman (IR) active.

$$
R(A_{1g})=\begin{pmatrix}
 a & d & 0\\
 d & a & 0 \\
 0 & 0 & b
\end{pmatrix}
;R(E_{g,1})=\begin{pmatrix}
 c & 0 & 0\\
 0 & -c & d \\
 0 & d & 0
\end{pmatrix}
$$

$$
and \ R(E_{g,2})=\begin{pmatrix}
 0 & -c & -d\\
 -c & 0 & 0 \\
 -d & 0 & 0
\end{pmatrix}
$$

For the space group \textit{P}2/\textit{m} with point group C$_{2h}$ ($2/m$):
$$\Gamma_{acoustic} = A_{u} + 2B_{u} $$
$$\Gamma_{optic} = 18A_{g} + 10A_{u} + 9B_{g} + 20B_{u}$$
In total, there are 60 vibrational modes, all of them are nondegenerate $A_{g}, B_{g}, A_{u} and B_{u}$ modes. Here, the optical vibration $10A_{u}$ + $20B_{u}$ is infrared (IR) active, while the optical mode $18A_{g}$ + $9B_{g}$ is Raman (R) active. The corresponding mode activity and symmetry at the $\Gamma$ point are shown in Table \ref{tabB2}.
$$
R(A_{g})=\begin{pmatrix}
 a & d & 0\\
 d & b & 0 \\
 0 & 0 & c
\end{pmatrix}
;R(B_{g})=\begin{pmatrix}
 0 & 0 & e\\
 0 & 0 & f \\
 e & f & 0
\end{pmatrix}
$$

\begin{table}
\caption{Wyckoff Positions (WP),Site Symmetry Group (SSG), and Mechanical Representation of every atoms in \textit{P}2/\textit{m} and \textit{R}$\bar{3}$\textit{m} space groups}
\begin{tabular}{cccccc}
\hline
\hline
SG                    & Atoms               & Coordinates & WP & SSG & Mechanical Rep.                 \\
\hline
\multirow{7}{*}{ \textit{P}2/\textit{m}} & \multirow{3}{*}{$A$} & (1/2,0,0)   & 1\textit{d}                & 2/\textit{m}                 & \multirow{2}{*}{$A_{u}+2B_{u}$}        \\
                      &                     & (0,0,1/2)   & 1\textit{c}                & 2/\textit{m}                 &                                \\
                      &                     &(\textit{x},1/2,\textit{z})   & 2\textit{n}                & \textit{m}                   & \multirow{5}{*}{$2A_{g}+A_{u}+B_{g}+2B_{u}$} \\
                      & \multirow{2}{*}{In} & (\textit{x},1/2,\textit{z})   & 2\textit{n}                & \textit{m}                   &                                \\
                      &                     & (\textit{x},0,\textit{z})     & 2\textit{m}                & \textit{m}                   &                                \\
                      & \multirow{2}{*}{As} & (\textit{x},1/2,\textit{z})   & 2\textit{n}                & \textit{m}                   &                                \\
                      &                     & (\textit{x},0,\textit{z})     & 2\textit{m}                & \textit{m}                   &                                \\
                      \hline
\multirow{3}{*}{\textit{R}$\bar{3}$\textit{m}} & $A$                  & (0,0,0)     & 3\textit{m}                & -3\textit{m}                 & $A_{2u}+E_{u}$                         \\
                      & In                  & (0,0,\textit{z})     & 6\textit{m}                & 3\textit{m}                  & \multirow{2}{*}{$A_{1g}+A_{2u}+E_{u}+E_{g}$} \\
                      & As                  & (0,0,\textit{z})     & 6\textit{m}                & 3\textit{m}                  &       \\
                      \hline
				\hline
\label{tabB1}
\end{tabular}
\end{table}

\begin{table*}
\centering
\caption{The irreducible representation of the transformation relation between the $D_{3d}$ point group and the $C_{2h}$ point group at the $\Gamma$-high symmetry point.}
\begin{tabular}{cc|cccccc}
\hline
\hline
$k$-vector                                     &                               & \multicolumn{6}{c}{$\Gamma$ (0, 0, 0)}                                                                                                          \\
\hline
\multirow{2}{*}{Relationsbetween the irreps} & \multicolumn{1}{c}{$D_{3d}$ ($-3m$)} & $A_{1g}$                  & $A_{1u}$                  & $A_{2g}$                  & $A_{2u}$                  & $E_{g}$                   & $E_{u}$                    \\
                                             & \multicolumn{1}{c}{$C_{2h}$ ($2/m$)} & $A_{g}$                   & $A_{u}$                   & $B_{g}$ & $B_{u}$ & $A_{g}$+$B_{g}$                & $A_{u}$+$B_{u}$      \\           
\hline
\hline
\end{tabular}
\label{tabB2}
\end{table*}

\begin{equation}
T=
\begin{pmatrix}
 -\frac{2}{3}  & -1 & -\frac{4}{3} & -\frac{1}{3}  \\
 \frac{2}{3}  & -1 & \frac{4}{3} & \frac{1}{3}  \\
 -\frac{1}{3}  & 0 & \frac{1}{3} & -\frac{1}{6}
\end{pmatrix}
\label{equ18}
\end{equation}

$$
GS=
\begin{pmatrix}
 0  & -1 & \frac{2}{3} & 0  \\
 0  & -1 & -\frac{2}{3} & 0  \\
 1  & 0 & -\frac{2}{3} & 0
\end{pmatrix} ; 
EAN=
\begin{pmatrix}
 -1  & 0 & -1 \\
 0  & 1 & 0  \\
 -1  & 0 & -2 \\
0 & 0 & 0
\end{pmatrix};
$$

$$
LC=
\begin{pmatrix}
 -\frac{2}{3}  & -1 & -\frac{4}{3}   \\
 \frac{2}{3}  & -1 & \frac{4}{3}  \\
 -\frac{1}{3}  & 0 & -\frac{1}{3} 
\end{pmatrix} ; 
EEN=
\begin{pmatrix}
 1  & 0 & 0 & \frac{1}{2}\\
 0  & 1 & 0  & 0 \\
 0  & 0 & 1  &0
\end{pmatrix}.
$$

The structural transformation is performed in Bilbao Crystallographic Server\cite{Ivantchev:ks0038}. The chain of transformation relations from the \textit{R}$\bar{3}$\textit{m} to the \textit{P}${2}$/\textit{m} structure includes three transformation matrices channels $(P_{i},p)$($i=1-6$) [see Eq. (\ref{equ19})]. These matrices achieve the symmetric operational transformation from the hexagonal to the monoclinic phase.
\begin{figure*}[!htbp]
\begin{equation}
\begin{aligned}
&(P_{1}|p)=
\begin{pmatrix}
 0  & 1 & 2/3 &| 0 \\
 0  & 0 & 4/3 &| 0\\
 1  & 0 & -2/3 &| 0
\end{pmatrix};
(P_{2}|p)=
\begin{pmatrix}
 -2/3  & 1 & 0 &| -1/6 \\
 -4/3  & 0 & 0 &| -1/3\\
 -4/3  & 0 & 1 &| -1/3
\end{pmatrix};
(P_{3}|p)=
\begin{pmatrix}
 0  & 0 & -4/3 &|-1/3 \\
 0  & 1 & -2/3 &| -1/6 \\
 1  & 0 & -2/3 &| -1/6
\end{pmatrix};\\
&(P_{4}|p)=
\begin{pmatrix}
 -4/3  & 0 & 4/3 &|0 \\
 -2/3  & 1 & 2/3 &| 0 \\
 1/3  & 0 & -4/3 &| 0
\end{pmatrix};
(P_{5}|p)=
\begin{pmatrix}
 -2/3  & -1 & 0 &|0 \\
 2/3  & -1 & 0 &| 0 \\
 -4/3  & 0 & 1 &| 0
\end{pmatrix};
(P_{6}|p)=
\begin{pmatrix}
 2/3  & -1 & -2/3 & | -1/6 \\
 -2/3  & -1 & 2/3 &| 1/6 \\
 1/3  & 0 & -4/3 & | -1/3 
\end{pmatrix}.
\end{aligned}
\label{equ19}
\end{equation}
\end{figure*}

\begin{table}
\caption{Symmetry operations of \textit{R}$\bar{3}$\textit{m} space group}
\begin{tabular}{ccccc}
\hline
\hline
SO & \multicolumn{1}{c}{Seitz symbols} & \multicolumn{1}{c}{(\textit{x},\textit{y},\textit{z}) form} & Matrix form &  \\
\hline
$\varepsilon$                     & \{1 $\vert$ 0 \}                     & \textit{x,y,z}                            &
$
\begin{pmatrix}
 1  & 0 & 0  \\
 0  & 1 & 0  \\
 0  & 0 & 1
\end{pmatrix}
$          &  \\
\textit{I}                     & \{-1 $\vert$ 0\}                    & \textit{-x,-y,-z}                         &
$
\begin{pmatrix}
 -1  & 0 & 0  \\
 0  & -1 & 0  \\
 0  & 0 & -1
\end{pmatrix}
$           &  \\
$\textit{C}_{2x}$                    & \{$2_{100} \vert$ 0\}                  & \textit{x-y,-y,-z}                        &
$
\begin{pmatrix}
 1  & -1 & 0  \\
 0  & -1 & 0  \\
 0  & 0 & -1
\end{pmatrix}
$           &  \\
$\textit{IC}_{2x}$                  & \{$\textit{m}_{100} \vert$ 0 \}                  & \textit{-x+y,y,z}                         &
$
\begin{pmatrix}
 -1  & 1 & 0  \\
 0  & 1 & 0  \\
 0  & 0 & 1
\end{pmatrix}
$           &  \\
$\textit{C}_{2y}$                   & \{$2_{010} \vert$ 0\}                  & \textit{-x,-x+y,-z}                       &
$
\begin{pmatrix}
 -1  & 0 & 0  \\
 -1  & 1 & 0  \\
 0  & 0 & -1
\end{pmatrix}
$           &  \\
$\textit{IC}_{2y}$                 & \{$m_{010} \vert$ 0\}                  & \textit{x,x-y,z}                          &
$
\begin{pmatrix}
 1  & 0 & 0  \\
 1  & -1 & 0  \\
 0  & 0 & 1
\end{pmatrix}
$           &  \\
$\textit{C}_{2xy}$                  & \{$2_{110} \vert$ 0\}                  & \textit{y,x,-z}                           &
$
\begin{pmatrix}
 0  & 1 & 0  \\
 1  & 0 & 0  \\
 0  & 0 & -1
\end{pmatrix}
$           &  \\
$\textit{IC}_{2xy}$                 & \{$m_{110} \vert$ 0\}                  & \textit{-y,-x,z}                          &
$
\begin{pmatrix}
 0  & -1 & 0  \\
 -1  & 0 & 0  \\
 0  & 0 & 1
\end{pmatrix}
$            & \\
\hline
\hline
\end{tabular}
\label{tabB3}
\end{table}

\begin{table}
\caption{Symmetry operations of \textit{P}${2}$/\textit{m} space group}
\begin{tabular}{ccccl}
\hline
\hline
SO & Seitz symbols & \multicolumn{1}{c}{(\textit{x},\textit{y},\textit{z}) form} & Matrix form &  \\
\hline
$\varepsilon$                  & \{1$\vert$0\}       & \textit{x,y,z}        &
$
\begin{pmatrix}
 1  & 0 & 0  \\
 0  & 1 & 0  \\
 0  & 0 & 1
\end{pmatrix}
$           &  \\
$C_{2y}$                 & \{2$_{010}$$\vert$0\}  & \textit{-x,y,-z}      &
$
\begin{pmatrix}
 -1  & 0 & 0  \\
 0  & 1 & 0  \\
 0  & 0 & -1
\end{pmatrix}
$                   &  \\
$I$                     & \{-1$\vert$0\}      & \textit{-x,-y,-z}     &
$
\begin{pmatrix}
 -1  & 0 & 0  \\
 0  & -1 & 0  \\
 0  & 0 & -1
\end{pmatrix}
$               &  \\
$IC_{2y}$                & \{m$_{010}$$\vert$0\}  & \textit{x,-y,z}       &
$
\begin{pmatrix}
 1  & 0 & 0  \\
 0  & -1 & 0  \\
 0  & 0 & 1
\end{pmatrix}
$              &  \\
\hline
\hline
\end{tabular}
\label{tabB4}
\end{table}

\subsection{\label{secB-3}Mechanical Properties Characterisation}
\subsubsection{\label{subB-3-1}Calculation of elastic constants}
\begin{figure}[!htbp]
	\begin{centering}
		\includegraphics[width=0.5\textwidth]{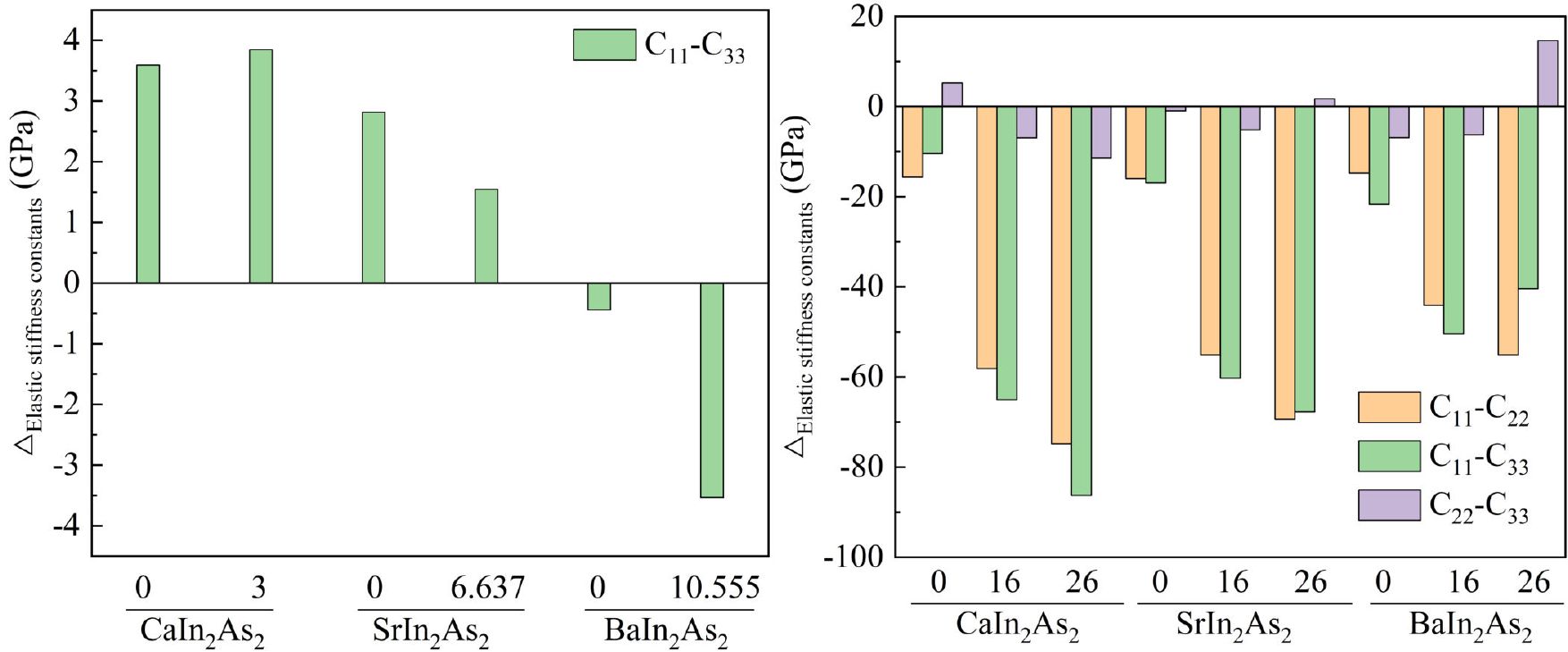}
		\par\end{centering}
	\centering{}\caption{The difference in elastic stiffness constants for the $A$In$_{2}$As$_{2}$ system belonging to the (a)\textit{P}6$_{3}$/\textit{mmc} (b) \textit{P}2/\textit{m} space group corresponds to the data in Table \ref{tabB5}.}
	\label{figB2}
\end{figure}

Stress and strain tend to change the elastic tensor information of the solid materials, so it is crucial to study the mechanical properties of materials under pressure, such as Young's modulus, shear modulus, $p$-wave modulus, Poisson's ratio, anisotropy index, Kleinman's parameter, Cauchy pressure, Pugh's ratio, and hardness information. Our calculated results under all pressures satisfy the criteria for mechanical stability in Appendix \ref{subA-2-1}, representing that all $A$In$_{2}$As$_{2}$ systems are mechanically stable. We calculated the elastic constants for the two phases at different pressures as shown in Table \ref{tabB5}. $C_{11}$, $C_{22}$ and $C_{33}$ denote the linear compression resistance along the $a$-, $b$- and $c$-axes, respectively. For hexagonal phase, $C_{11}$=$C_{22}\neq C_{33}$ and $C_{33}$ are smaller than $C_{11}$ for all systems except BaIn$_{2}$As$_{2}$, indicating that the $c$-axis is more compressible than the $a$-axis and $b$-axis, which also reflects the weaker chemical bonding in the $c$-axis than the $a$-axis and $b$-axis. In contrast, $C_{33}$ is larger than $C_{11}$ in BaIn$_{2}$As$_{2}$ of the hexagonal phase resulting in the $c$-axis being more incompressible than the $a$($b$) axis, indicating that the $c$-directional chemical bonding of BaIn$_{2}$As$_{2}$ is more stable than the $a$($b$)-directional. This easily compressible direction evaluates the maximum probability direction of the structure phase transition. Due to the weaker bonding in the $ab$ plane, the octahedral layers in Fig. \ref{fig1}(a) are more easily deformed within the layers than between them. This anomalous behavior of the hexagonal phase BaIn$_{2}$As$_{2}$ compared to CaIn$_{2}$As$_{2}$, SrIn$_{2}$As$_{2}$ perfectly explains the previous experimental result for their different structural phases. As can be seen from Fig. \ref{figB2}(a), the bonding strengths differences between the $a$($b$) axis and $c$ axis of CaIn$_{2}$As$_{2}$ and SrIn$_{2}$As$_{2}$ are positive, while the bonding in the in-plane ($C_{11}$ and $C_{22}$) of BaIn$_{2}$As$_{2}$ is weaker than that in the out-plane ($C_{33}$) (about -3.5 GPa). This result implies that CaIn$_{2}$As$_{2}$ and SrIn$_{2}$As$_{2}$ are prone to structural phase transitions in the $c$-direction, while BaIn$_{2}$As$_{2}$ is prone to structural phase transitions in the in-plane. 
For the monoclinic phase [see Table \ref{tabB5} and Fig. \ref{figB2}(a)], satisfying $C_{11} \neq C_{22} \neq C_{33}$, $C_{11}$ is smaller than $C_{22}$ and $C_{33}$ for all systems, and the difference $\Delta_{C_{11}-C_{ij}} (i=j=2,3)$ becomes more significant as the pressure is applied except for BaIn$_{2}$As$_{2}$ which becomes smaller under 26 GPa. Without pressure, $C_{22}$ of CaIn$_{2}$As$_{2}$ is maximum while $C_{33}$ of SrIn$_{2}$As$_{2}$ and BaIn$_{2}$As$_{2}$ is maximum. With applying pressure, $C_{22}$ and $C_{33}$ compete, CaIn$_{2}$As$_{2}$ becomes maximum at 16 GPa for $C_{33}$ while SrIn$_{2}$As$_{2}$ and BaIn$_{2}$As$_{2}$ reverse to the maximum at 26 GPa for $C_{22}$. In conclusion, the monoclinic phase of $A$In$_{2}$As$_{2}$, especially after applying pressure, has weak bonding in the $a$-axis, and it is relatively easy to peel in that direction.

\subsubsection{\label{subB-3-2}Elastic modulus analysis}

The bulk modulus $B$ is a physical measure of the material's ability to resist compression: the more significant the $B$, the more excellent the resistance to compression and the smaller the compressibility.
The $B$ of Hill approximation is related to $B_{V}$ and $B_{R}$, and the $B$ of both two phases can be explicitly calculated by Eqs. (\ref{equ7}),(\ref{equ9}), (\ref{equ11}), (\ref{equ13}). As shown in Table \ref{tabB5}, the $B$ of the monoclinic phase is generally smaller and more compressible than the hexagonal phase, which is related to the low symmetry structure of the monoclinic. The bulk modulus of the hexagonal phase obtained from the elastic constants remarkably agrees with the fit of the B-M equation reported in our previous work\cite{D2CP01764D}. The bulk moduli of the present paper (Ref. \cite{D2CP01764D}) are 45.779 GPa (46.3 GPa), 43.437 GPa (43.8 GPa), and 41.077 GPa (41.7 GPa) for the hexagonal phases CaIn$_{2}$As$_{2}$, SrIn$_{2}$As$_{2}$, and BaIn$_{2}$As$_{2}$, respectively. Our results also show that pressure can effectively enhance the resistance to compression of both two phases, which can be explained by the decrease in lattice parameters after compression (see Figs. \ref{fig3}(a)-\ref{fig3}(e) and Fig. S2 in Ref. \cite{D2CP01764D}).

The shear modulus $G$ reflects the ratio of stress to strain under shear deformation. The larger the $G$, the greater the resistance to shear deformation. $G$ can also be calculated from Eqs. (\ref{equ8}), (\ref{equ10}), (\ref{equ12}), (\ref{equ14}). The relationship between the two structural phases of $G$ and the trend of change under pressure is similar to that of $B$. The applied pressure can enhance the shear deformation resistance of most of the systems. However, it reduced again that the shear deformation resistance of CaIn$_{2}$As$_{2}$ at 26 GPa and BaIn$_{2}$As$_{2}$ (both two phases) under higher pressures. Without pressure, $B$ and $G$ of both space groups decrease as the atomic number of $A$ increases. 

Young's modulus $E$ is an important index to characterize the stiffness of solid materials reflecting the system's resistance to elastic deformation. Poisson's ratio $v$ reflects the stability of the solid against shear deformation. They can be calculated from $B$ and $G$ by Eq. (\ref{equ6}). Also given by Table \ref{tabB5}, the variation regular of E under pressure is consistent with $G$ and $B$ for the \textit{P}2/\textit{m} phase and \textit{P}6$_{3}$/\textit{mmc} phase, respectively. Poisson's ratio $v$ is stable at -1$ \sim $0.5 under linear elastic shear deformation. Based on the data in Table \ref{tabB5}, we can quickly determine that $v$ is positive and within the stability range, again proving that all systems are mechanically stable.

\begin{table*}
\centering
\caption{Elastic stiffness constants $C_{ij}$(GPa), bulk modulus $B$ (GPa), shear modulus $G$ (GPa), Young's modulus $E$ (GPa), and Poisson's ratio $v$ from different space group: monoclinic phases ($\textit{P}2/\textit{m}$) and hexagons ($\textit{P}6_{3}/\textit{mmc}$) for $A$In$_{2}$As$_{2}$ under different pressures.}
\resizebox{\textwidth}{!}{
\begin{tabular}{cccccccccccccccccccc}
\hline
    \hline
   Phase     &    Material      &   Pressure (GPa)     & $C_{11}$     & $C_{12}$      & $C_{13}$     & $C_{22}$     & $C_{23}$    & $C_{33}$      & $C_{44}$      & $C_{15}$    & $C_{25}$    & $C_{35}$   & $C_{55}$    & $C_{64}$   & $C_{66}$      & $B$       & $G$      & $E$      & $v$     \\
        \hline
$\textit{P}2/\textit{m}$    & CaIn$_{2}$As$_{2}$ & 0      & 74.652   & 32.609   & 23.923   & 90.334  & 16.44  & 85.140    & 20.326   & 1.384  & 1.644  & 1.356 & 23.806 & 2.523 & 35.872   & 43.856  & 27.09  & 67.386 & 0.244 \\
        &          & 16     & 87.617   & 69.327   & 56.958   & 145.725 & 49.167 & 152.6681 & 32.712   & $-$1.610  & 2.573  & 8.960  & 33.992 & 8.477 & 45.378   & 79.655  & 33.877 & 88.987 & 0.314 \\
        &          & 26     & 87.971   & 88.590    & 92.947   & 162.791 & 90.496 & 174.268  & 47.198   & $-$5.017 & $-$1.149 & 6.888 & 42.145 & 5.176 & 52.504   & 97.323  & 32.028 & 86.503 & 0.353 \\
        & SrIn$_{2}$As$_{2}$ & 0      & 70.374   & 32.771   & 20.298   & 86.344  & 12.462 & 87.300     & 19.110    & 1.273  & 2.209  & 2.041 & 22.938 & 3.166 & 38.071   & 41.525  & 26.743 & 66.033 & 0.235 \\
        &          & 16     & 89.850    & 75.557   & 49.900     & 144.961 & 48.398 & 150.152  & 32.796   & $-$1.868 & 2.524  & 8.731 & 31.823 & 6.686 & 49.650    & 79.198  & 33.985 & 89.161 & 0.313 \\
        &          & 26     & 99.850    & 101.189  & 75.526   & 169.273 & 88.088 & 167.593  & 47.132   & $-$4.574 & 1.583  & 7.289 & 37.002 & 4.308 & 52.987   & 100.894 & 34.213 & 92.168 & 0.348 \\
        & BaIn$_{2}$As$_{2}$ & 0      & 64.772   & 32.308   & 17.798   & 79.566  & 10.205 & 86.469   & 17.118   & 1.027  & 2.697  & 2.239 & 20.904 & 3.411 & 38.577   & 38.900    & 25.110  & 61.958 & 0.235 \\
        &          & 16     & 93.718   & 81.344   & 47.987   & 137.817 & 51.239 & 144.157  & 29.619   & $-$3.343 & 2.288  & 6.966 & 27.790  & 4.168 & 49.850    & 80.026  & 31.199 & 82.778 & 0.328 \\
        &          & 26     & 106.797  & 109.860   & 68.359   & 161.884 & 95.283 & 147.284  & 41.861   & $-$4.215 & 3.589  & 2.877 & 29.211 & 0.906 & 50.607   & 99.001  & 28.757 & 78.603 & 0.369 \\
$\textit{P}6_{3}/\textit{mmc}$ & CaIn$_{2}$As$_{2}$ & 0      & 88.379   & 24.956   & 25.243   &   -    &   -   & 84.788   & 25.879   &    -    &    -    &   -    &     -   &    -   & 31.712   & 45.779  & 29.093 & 72.022 & 0.238 \\
        &          & 3      & 101.826  & 32.380    & 34.540    &    -     &    -    & 97.980    & 26.634   &   -     &     -   &   -    &   -     &    -   & 34.723   & 56.027  & 30.820  & 78.133 & 0.268 \\
        & SrIn$_{2}$As$_{2}$ & 0      & 82.964   & 23.707   & 24.423   &   -      &   -     & 80.149   & 26.357   &   -     &    -    &   -    &   -     &    -   & 29.629   & 43.437  & 28.072 & 69.289 & 0.234 \\
        &          & 6.637  & 107.796  & 39.817   & 45.241   &     -    &    -    & 106.251  & 25.065   &     -   &   -     &    -   &   -     &  -     & 33.989   & 64.672  & 29.423 & 76.645 & 0.302 \\
        & BaIn$_{2}$As$_{2}$ & 0      & 75.246   & 23.301   & 24.091   &     -    &     -   & 75.69    & 25.760    &   -     &   -     &    -   &    -    &   -    & 25.973   & 41.077  & 25.626 & 63.643 & 0.242 \\
        &        & 10.555 & 112.472 & 54.953 & 55.710 &    -     &    -    & 116.004 & 17.994 &   -     &    -    &     -  &    -    &   -    & 28.759 & 74.792  & 24.098 & 65.278 & 0.355 \\
         \hline
         \hline
\end{tabular}}
\label{tabB5}
\end{table*}

\begin{table*}
\centering
\caption{The reciprocal of Pugh's ratio $B/G$, Kleinman's parameter $\xi$, Cauchy's pressure $P_{C}$ (GPa), chemical bond type, universal anisotropy index ($A_{u}$), and Log-Euclidean anisotropy ($A_{L}$) from different space group: monoclinic phases ($\textit{P}2/\textit{m}$) and hexagons ($\textit{P}6_{3}/\textit{mmc}$) for $A$In$_{2}$As$_{2}$ under different pressures.}
\resizebox{\textwidth}{!}{
\begin{tabular}{cccccccccccccccccccccc
}
\hline
\hline
Phase   & Material & Pressure (GPa)  & B/G &  $\xi$ & $P_{C}^{a}$ (Chemical bond) & $P_{C}^{c}$ (Chemical bond) & $A_{u}$ & $A_{L}$ \\
\hline
\multirow{9}*{$P$2/$m$}    & CaIn$_{2}$As$_{2}$ & 0              & 1.619 (brittle)           & 0.73                 & 12.3 (MLB)  &   - & 0.315                          & 0.134                          \\
        &          & 16             & 2.351 (ductile)           & 1.35                 & 36.6 (MLB)             & -              & 0.910                          & 0.357                          \\
        &          & 26             & 3.04  (ductile)          & 1.81                 & 41.4 (MLB)             & -             & 2.883                          & 0.973                          \\
        & SrIn$_{2}$As$_{2}$ & 0              & 1.553 (brittle)           & 0.78                 & 13.7 (MLB)             & -              & 0.491                          & 0.207                          \\
        &          & 16             & 2.330 (ductile)           & 1.45                 & 42.8 (MLB)            & -              & 1.022                          & 0.398                          \\
        &          & 26             & 2.949 (ductile)           & 1.83                 & 54.1 (MLB)             & -             & 1.753                          & 0.639                          \\
        & BaIn$_{2}$As$_{2}$ & 0              & 1.549 (brittle)           & 0.83                 & 15.2 (MLB)             & -              & 0.701                          & 0.291                          \\
        &          & 16             & 2.565 (ductile)           & 1.51                 & 51.7 (MLB)             & -             & 1.204                          & 0.468                          \\
        &          & 26             & 3.443 (ductile)           & 1.87                 & 68.0 (MLB)            & -              & 2.255                          & 0.794                          \\
	\multirow{6}*{$P$6$_{3}$/$mmc$} & CaIn$_{2}$As$_{2}$ & 0              & 1.574 (brittle)         & 0.50                 & $-$0.6 (CLB)             & $-$6.8 (CLB)           & 0.044                          & 0.020                          \\
        &          & 3              & 1.818 (ductile)           & 0.56                 & 7.9 (MLB)              & $-$2.3 (CLB)             & 0.072                          & 0.032                          \\
        & SrIn$_{2}$As$_{2}$ & 0              & 1.547 (brittle)           & 0.51                 & $-$1.9 (CLB)             & $-$5.9 (CLB)          & 0.014                          & 0.006                          \\
        &          & 6.637          & 2.198 (ductile)           & 0.63                 & 20.2 (MLB)             & 5.8 (MLB)              & 0.094                          & 0.041                          \\
        & BaIn$_{2}$As$_{2}$ & 0              & 1.603(brittle)            & 0.55                 & $-$1.7 (CLB)            & $-$2.7 (CLB)          & 0.000                          & 0.000                          \\
        &          & 10.555         & 3.104 (ductile)           & 0.81                 & 37.7 (MLB)             & 26.2 (MLB)              & 0.275                          & 0.120
        \\
\hline
\hline
\end{tabular}}
\label{tabB6}
\end{table*}

\subsubsection{\label{subB-3-3}Pressure affects crystal anisotropy results}
In order to visualize the effect of pressure on each elastic modulus, we calculated their 2D projections in a specific plane (see Figs. S6-S10\cite{SI}). The hexagonal phase (see Figs. S6 and S7\cite{SI}) tends to be more isotropic than the monoclinic phase due to the higher symmetry and the neater octahedral lattice. Especially with the most significant isotropy for each BaIn$_{2}$As$_{2}$ mechanics without pressure (compare with Figs. S6 and S10 within the SM \cite{SI}). When the pressure modulates $A$In$_{2}$As$_{2}$ as a zero-bandgap solid material, the linear compression maintains isotropy. Since the pressure values of induced zero bandgaps for CaIn$_{2}$As$_{2}$ and SrIn$_{2}$As$_{2}$ are weak, the variation of each mechanical quantity is not significantly different from that under no pressure. However, we can see that a pressure of 10.555 GPa will induce the change of $G$, $E$, $v$ of BaIn$_{2}$As$_{2}$ from isotropic to anisotropic. For CaIn$_{2}$As$_{2}$ in the monoclinic phase (see Fig. S8 within the SM\cite{SI}), the following conclusions can be drawn as the pressure increases: 1. The anisotropy of the $G$-minimum positive (green curve) and $E$-maximum positive increases, especially in the $xy$(001) and $xz$(010) planes. This phenomenon is because the difference between $C_{22}$ ($C_{33}$) and $C_{11}$ increases sharply under pressure. 2. The maximum positive value of linear compression changes from almost isotropic to polarized in the $x$-direction ($a$-axis). The maximum positive value in the $yz$-plane disappears due to the pressure effect. In contrast, the minimum negative value polarization along the $z$-direction ($c$-axis) appears under 26 GPa pressure (red curve). 3. At 26 GPa, the minimum NPR phenomenon occurs (red curve in Fig. S8 within the SM\cite{SI}). For SrIn$_{2}$As$_{2}$ and BaIn$_{2}$As$_{2}$ in the monoclinic phase of Figs. S9 and S10 within the SM\cite{SI}, the more obvious difference is that the pressure-induced linear compression at 26 GPa has a minimum negative value along the $y$-direction ($b$-axis), but not $z$-direction. Second, the maximum value of linear compression and the minimum positive value of $v$  (green curve) are observed in the $yz$ plane under 26 GPa, which are not visible in CaIn$_{2}$As$_{2}$. Also, the anisotropy of the maximum positive value of $v$ (blue curve) for the CaIn$_{2}$As$_{2}$ and SrIn$_{2}$As$_{2}$ regimes at 26 GPa is weaker than that of BaIn$_{2}$As$_{2}$ in $yz$ plane. 

Furthermore, we calculate the 3D space-dependent mechanical quantities ($G$, $E$, $B$ and $v$) for the two phases as BaIn$_{2}$As$_{2}$ (see Figs. S1-S5\cite{SI}). It can be visualized that the $A$In$_{2}$As$_{2}$ of the intrinsic hexagonal phase is indeed highly isotropic, while the monoclinic phase exhibits anisotropy.

\subsubsection{\label{subB-3-4}Chemical bonding, brittleness and hardness prediction}

The Pugh's ratio ($G/B$) or $B/G$ ratio defines the ductility or brittleness of a solid. With $B/G$= 1.75 as the threshold value, a material with $B/G$\textgreater 1.75 is considered ductile, while the opposite is considered brittle. From Table\ref{tabB6}, we can find that the $A$In$_{2}$As$_{2}$ system without pressure behaves as brittle, while $B/G$ increases and transforms into ductile after applying pressure. 
The $\xi$ parameter evaluates whether the material is bending-dominated or stretch-dominated. When the value is close to 0, the system is bending-dominated, and close to 1, it is tensile-dominated. As seen in Table\ref{tabB6}, the monoclinic phase has a larger $\xi$ than the hexagonal phase, and the larger the $A$ atomic number, the larger the $\xi$. Moreover, all the $A$In$_{2}$As$_{2}$ systems we consider, whether pressurized or not, exhibit stretching dominance.
As demonstrated in Table \ref{tabB6}, Cauchy's pressure $P_C$ can effectively assess the type of chemical bonding. The monoclinic phase $A$In$_{2}$As$_{2}$ tends to bond in a metallic manner MLB, and the strength of this bonding is proportional to the ionization energy of the $A$ ion. Moreover, the pressure favors the enhancement of the metallic character of the system. The hexagonal phase of $A$In$_{2}$As$_{2}$ has both Cauchy's pressures ($P_{C}^{a}$ and $P_{C}^{c}$) negative in the absence of pressure, indicating a tendency to form covalent bonds CLB. In addition, the pressure will reverse the sign of Cauchy's pressure, and the bonding style changes to a metallic bonding-dominated situation.

\begin{figure}[!htbp]
	\begin{centering} 
		\includegraphics[width=0.5\textwidth]{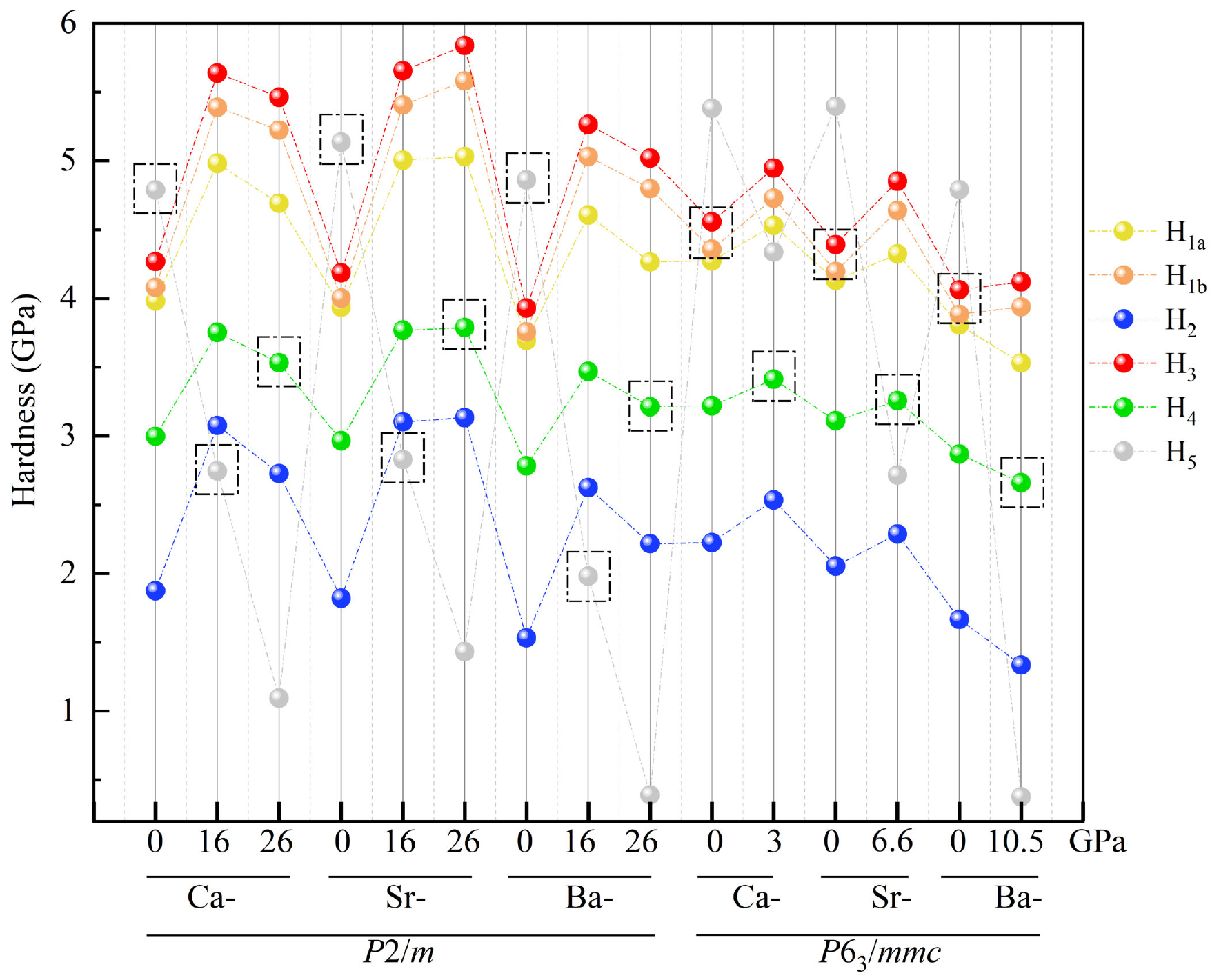}
		\par\end{centering}
	\centering{}\caption{Comparison of hardness parameters of $A$In$_{2}$As$_{2}$ hexagonal phase (\textit{P}${6}_{3}$/\textit{mmc}) and monoclinic phase (\textit{P}${2}$/\textit{m}) under different pressures.}
	\label{figB3}
\end{figure}

\begin{figure}[!htbp]
	\begin{centering}
		\includegraphics[width=0.5\textwidth]{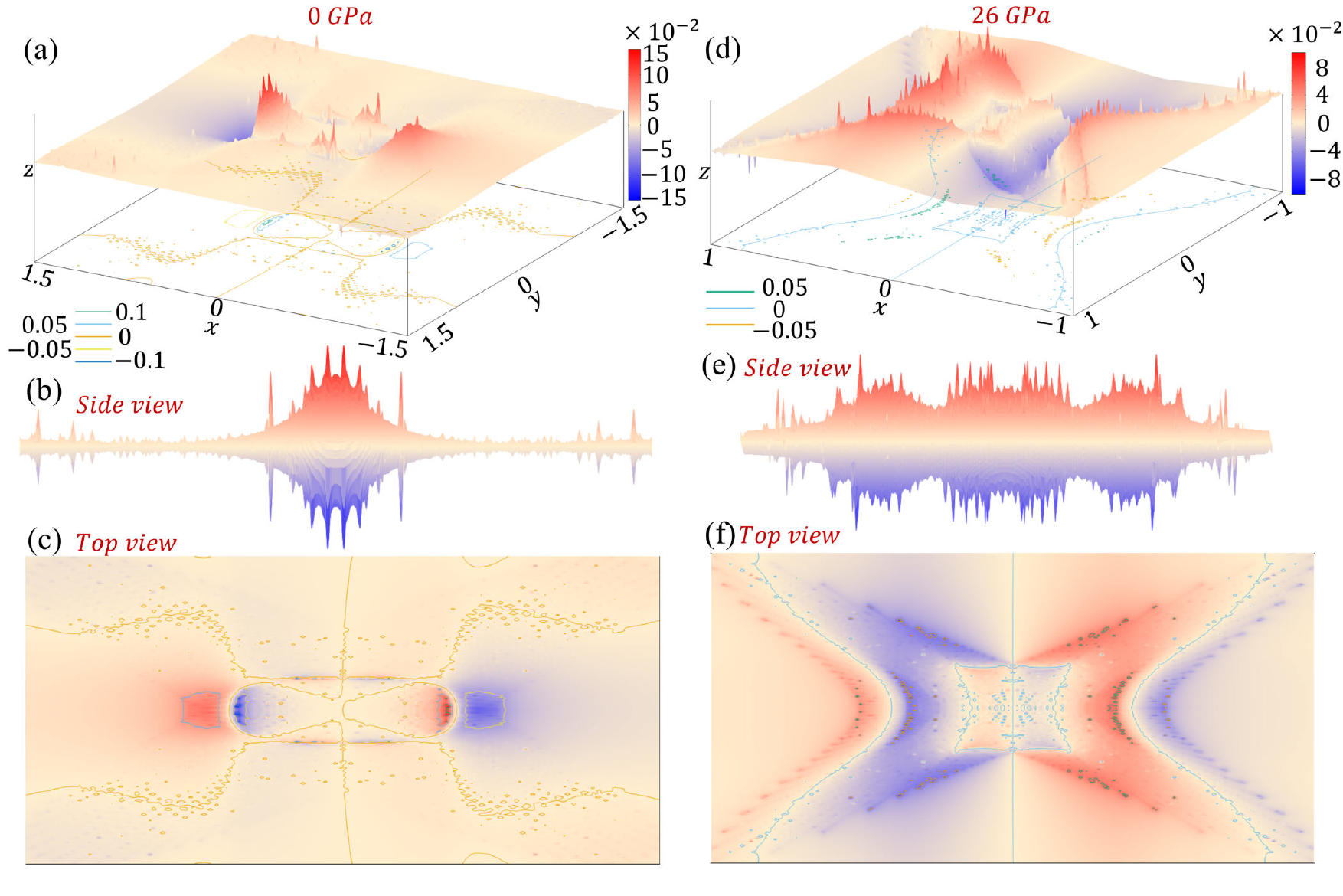}
		\par\end{centering}
	\centering{}\caption{Hardness distribution of the BaIn$_{2}$As$_{2}$ monoclinic phase structure under 0 and 26 GPa pressures.}
	\label{figB4}
\end{figure}

Hardness can adequately describe the mechanical behavior of solids and is one of the critical factors in practical production processes. We evaluate the hardness of $A$In$_{2}$As$_{2}$ in different states according to the six semi-empirical formulas of Eq. (\ref{equ17}) and the judgment guide of Table \ref{tabA1}.  The Vickers hardness is calculated from $H_{1b}$ or $H_{3}$ for the hexagonal phase of $A$In$_{2}$As$_{2}$ with $P6_3mmc$ space group, semiconductors at 0 GPa (0  \textless $E_{g}$  \textless 2 eV). As shown in the orange and red curves in Fig. \ref{figB3}, they are higher than other calculations. When applying a pressure that induces a zero band gap ($E_{g}$=0), the crystal hardness tends to be expressed by $H_{4}$, with a reduced hardness (green curve). For the monoclinic phase of the general case, $A$In$_{2}$As$_{2}$ is a semiconductor at 0 GPa and 16 GPa and becomes metallic at 26 GPa. Again, from the results of Table \ref{tabA1} and Fig. \ref{figB3}, we know that the $H_{5}$ equation can represent the hardness of the system at 0 GPa and 16 GPa, while $H_{4}$ describes the hardness of the system at 26 GPa. In the absence of pressure, the monoclinic phase of $A$In$_{2}$As$_{2}$ has the maximum hardness (indicated by the grey curve). With a pressure of 16 GPa, the hardness is still expressed by $H_{5}$, but the hardness decreases by almost half, especially for BaIn$_{2}$As$_{2}$. At a pressure of 26 GPa, the hardness of the system increases again and is expressed by $H_{4}$. We suggest that the change in hardness may have a necessary relationship to the structural phase transition. Overall, our predicted hardness of the $A$In$_{2}$As$_{2}$ material is not high, well below the experimental 96 GPa for diamond\cite{KLEIN1993918}, but close to that of ZnO (7.2 GPa)\cite{83JIANG20112287}, which is also a hexagonal phase. 

In order to observe more comprehensively the effect of pressure on the overall hardness of the crystal, we calculated 3D hardness distributions for BaIn$_{2}$As$_{2}$ as an example (see Figs. \ref{figB4}).  The left and right columns of Fig. \ref{figB4} show the hardness distributions of monoclinic phase BaIn$_{2}$As$_{2}$ at 0 GPa and 26 GPa, respectively. The hardness distribution is symmetric about the $x$-axis when no pressure is applied [see Fig. \ref{figB4}(a)], and the symmetry of the hardness distribution is broken when 26 GPa pressure is applied [see Fig. \ref{figB4}(d)]. Such an asymmetric transition can be observed more clearly in the side ($yz$-plane) of Figs. \ref{figB4}(b) and \ref{figB4}(e). The range of coordinates and the intensity of the contours in Figs. \ref{figB4}(a) and \ref{figB4}(d) allow determining that the maximum hardness of the 0 GPa crystal is higher than 26 GPa, consistent with the hardness relationship predicted quantitatively earlier. In addition, it can be found that the pressure application induces a shift in the hardness distribution of BaIn$_{2}$As$_{2}$ from clustering in the center of the crystal to dispersion in the $y$-direction. Figures \ref{figB4}(c) and \ref{figB4}(f) show the projection of hardness in the $xy$ plane. A comparison of the localized peak in Fig. \ref{figB4}(c) with the "fishtail" hardness relationship in Fig. \ref{figB4}(f) shows that pressure does weaken the hardness localization. 

\subsubsection{\label{subB-3-5}Analysis of the degree of $\Delta{F}-T$ linear correlation}
We are concerned that at high temperatures ($\ge$ 2500 K), the free energy difference of the pressure-absent $A$In$_{2}$As$_{2}$ system exhibits almost a linear decrease with temperature and has different slopes (see the enlarged figure in the upper right of Fig. S15(b) within the SM\cite{SI}). The temperature dependence of the energy difference between the two structural phases changes from parabolic to linear, and this monotonically decreasing relationship indicates that the phase transition from hexagonal to monoclinic has fully realised at ultrahigh temperatures. The degree of tilt of the curve depends on the $A$ atomic radius size. To better illustrate the linearity, we performed an error analysis of the slopes over the full range (0-3000 K) with the high-temperature linear slopes of the three systems, and the results are shown in Fig. S16 within the SM\cite{SI}. From the marked slope errors of 5$\%$ of the temperature values (2400 K, 2450 K, and 2420 K for CaIn$_{2}$As$_{2}$, SrIn$_{2}$As$_{2}$ and BaIn$_{2}$As$_{2}$, respectively), it can be seen that, within the error tolerance, the three systems show a linear slope after temperatures above 2500 K. The slope is linear for all three systems after the temperature above 2500 K within the error tolerance. 

\end{appendices}
\clearpage
\end{CJK}
\bibliography{Refs}
%

\end{document}